\documentclass[aps,phyrev, reprint,amsmath,amssymb,superscriptaddress
]{revtex4-2}
\usepackage[caption=false]{subfig}
\usepackage{graphicx}
\usepackage[english]{babel}
\usepackage{longtable} 
\usepackage{multirow}
\usepackage{bm}
\usepackage{textcomp}
\usepackage[usenames]{color}
\usepackage[usenames,dvipsnames]{xcolor}
\usepackage{mathptmx}
\usepackage[utf8]{inputenc}
\usepackage{hyperref}
\hypersetup{%
    bookmarks=true,%
    unicode=true,%
    pdfsubject={paper},%
    linktocpage,%
    colorlinks,%
    citecolor=blue,%
    filecolor=black,%
    linkcolor=blue,%
    urlcolor=blue
}

\def\mpdlr{\affiliation{%
Institut f\"ur Materialphysik im Weltraum, Deutsches Zentrum f\"ur Luft- und Raumfahrt (DLR), 51170 K\"oln, Germany}}
\def\frm{\affiliation{%
Heinz Maier-Leibnitz-Zentrum, Technische Universit\"at M\"unchen, Lichtenbergstra\ss e 1, 85748 Garching}}
\def\ill{\affiliation{%
Institut Laue-Langevin (ILL), 71 avenue des Martyrs, 38042 Grenoble, France}}
\def\simap{\affiliation{%
Université Grenoble Alpes, CNRS, Grenoble INP, SIMaP, F-38000 Grenoble, France}}

\begin{document}
\title{Impact of local structure on melt dynamics in Cu-Ti alloys: Insights from ab-initio molecular dynamics simulations}
\author{Lucas P. Kreuzer} 
\email{lucas.kreuzer@frm2.tum.de}
\frm
\mpdlr
\author{Fan Yang} 
\mpdlr


\author{Andreas Meyer} 
\mpdlr
\ill

\author{No\"el Jakse}
\simap

\date{\today}
\begin{abstract}
First-principle based molecular-dynamics simulations have been performed for binary Cu$_x$Ti$_{1-x}$ (x = 0.31, 0.50, and 0.76) alloys to investigate the relationship between local structure and dynamical properties in the liquid and undercooled melt. The undercooled melts show a pronounced short-range order, majorly a five-fold symmetry (FFS) around the Cu atoms, which competes with bcc ordering. This complex SRO is also reflected in the partial coordination numbers, where mainly a Z12 coordination is present around Cu, which corresponds to an icosahedral ordering. Higher coordination numbers were obtained for Ti compatible with Frank-Kasper polyhedra. The increasing Frank-Kasper polyhedra coordination scenario around Ti impacts on the interatomic distances of Ti atoms, which increase with increasing Ti content. The Cu$_{50}$Ti$_{50}$ composition exhibits the highest FFS ordering and amount of Frank-Kasper polyhedra, which explains the slowest melt dynamics, found experimentally and in simulations for this composition. Thus, our results suggest that the high undercooling degree and glass-forming ability of binary CuTi alloys, originates from the high complexity of the local structure rather than due to the preferred formation of Cu-Ti pairs, as Cu-Ti interactions were found to be weak.
\end{abstract}

\maketitle
\section{Introduction}
Bulk metallic glasses (BMGs) feature an amorphous structure in their solid state, which gives them superior mechanical properties as compared to their crystalline counterparts. Besides complex multi-component alloy systems, also relatively simple binary systems are known to show excellent glass-forming abilities (GFA) \cite{Xu_2004, Wang_2004, Xia_2006a, Xia_2006b, Miracle_2010}. Even for binary systems the underlying mechanisms responsible for a good GFA are still not fully understood on atomic length and time scales \cite{Zhang2001, Lee2008, Wu2008, Amore2011a, Gargarella2015}. Using them as model systems to understand structural and dynamical properties, especially in the undercooled region, where nucleation and growth \cite{Shang2023} or glass formation \cite{Cao2024} take place, is of utmost importance. 

Among the most well-known and widely studied binary glass-forming alloys are Zr-based systems, e.g., Cu-Zr and Ni-Zr. Over the last years, numerous experimental and computational studies were reported on the relationship between structure and dynamics in the melt in order to explain the observed high GFA \cite{Wang_2004, Mei-Bo2004, Jakse_2008a,Yang2014, Jakse2013b}. It has been found that the composition-dependent dynamics in Zr-Cu can be understood by the packing density, i.e., a more densely packed melt, results in slower melt dynamics \cite{Yang2014}. When replacing Cu with Ni, which has a similar atomic radius, the melt dynamics become slower, but the GFA decreases \cite{Holland-Moritz2009, Holland-Moritz_2012}. This is surprising, since sluggish dynamics usually promote glass formation \cite{Herlach_2007}. A higher nucleation barrier for the Zr-Cu system due to a pronounced icosahedral and chemical short-range order (SRO), especially on the Cu-rich side of the phase diagram, might be an explanation \cite{Jakse_2008a}. 
When substituting Zr by Hf (both are chemically similar, as they have approximately the same atomic radius and are located in the same main group of the periodic table), the melt dynamics become slower, which is attributed to a locally higher packing density \cite{Nowak2017a}. Furthermore, for Zr-Ni melts a decoupling of the Ni diffusivity from the Zr diffusivity can be observed in the Ni-rich compositions, which can be attributed to the saturation of chemical interactions between Zr and Ni. Thus, besides packing and dynamical arguments, also chemical interactions seem to play a crucial role for the structure-dynamics relationship in the melt, and thus for the GFA.

Substitution of Zr by Ti, which stands above Zr in the 4$^\text{th}$ element group, underlines this presumption. In a recent study by us, we found a mixing behavior of the molar volume close to an ideal solution, resulting in an almost constant packing fraction for the binary Cu-Ti system \cite{Kreuzer_2024}. The obtained values of the packing fraction are similar to that of the Cu-Zr system, but slightly lower than that of the Ni-Zr system \cite{Yang2014, Holland-Moritz2009}. The melt viscosity of Cu-Ti features a non-monotonous trend with increasing Ti concentration and exhibits the largest values around an intermediate Ti concentration of 50 at\%. The slowdown of the dynamics was referred to chemical interactions between Cu and Ti, which form preferred neighboring pairs. Indeed, a chemical SRO was found, however it is less pronounced as e.g., for Ni-Zr and Ni-Hf, which might explain the overall faster atomic dynamics in the Cu-Ti system as compared to above mentioned Ni-based alloys \cite{Holland-Moritz2009,Nowak2017a}. These differences in average packing fraction and liquid viscosity point clearly to composition- and concentration-dependent interactions between the alloy components within these systems.

Especially, the large differences between the Cu-Zr and Cu-Ti systems, regarding the structure-dynamics relationship, despite a similar electronic configuration and atomic radius (r$_{Zr}$ = 1.450 \AA; r$_{Ti}$ = 1.324 \AA), cannot be explained, yet. Hence, information about how chemical interactions control the liquid dynamics over a broad temperature range and as a function of composition are needed e.g., by accessing partial structure factors, particularly $S_\text{CC}(q)$. However, measuring partial structure factors is challenging, and so far only the $S_\text{CC}(q)$ for Cu$_{31}$Ti$_{69}$ at 1000 K has been obtained experimentally \cite{Kreuzer_2024}. An elegant solution to this is the use of simulations, validated on the existing experimental data of the Cu-Ti system, and provide access to structure and dynamics over a broad composition range. Thereby, the role of chemical interactions for the liquid dynamics can be resolved, which contributes to an understanding of how the structure-dynamics relationship is (or is not) correlated with the GFA, and if such a relationship is universal for other (binary) alloys as well. Besides a fundamental interest, answering these questions is also of relevance for industrial processes. Ti is cheaper and lighter than Zr, which would enable new application fields, especially in the light-weight sector e.g., as alloys for aerospace or automotive applications.

The present work aims at resolving the structure-dynamics relationship and shed new light on recent experimental studies on the Cu-Ti system. \textit{Ab-initio} molecular dynamics (AIMD) simulations within the density functional theory (DFT) were performed to elucidate the structural and dynamic properties of liquid and undercooled Cu-Ti alloys \cite{Jakse_2008b, Woodward_2010, Jakse_2016, Pasturel_2016, Li_2023}. A compositional range spanning from 24 to 69 at.\% Ti, specifically targeting the compositions of Cu$_{76}$Ti$_{24}$, Cu$_{50}$Ti$_{50}$, and Cu$_{31}$Ti$_{69}$, which were also investigated experimentally, across a broad temperature range from 1000 to 1700 K, including also former results from pure Cu and Ti. The simulations provide detailed insights into the local structure, partial interatomic particle distances, coordination number, composition-dependent diffusion coefficients, and present symmetries in the melt and undercooled liquid, which are experimentally not accessible, but needed for a fully comprehensive understanding of the structure-dynamics relationship and its impact on the GFA. 

In this study, a pronounced five-fold symmetry (FFS) in the undercooled melt is observed, particularly around Cu atoms, competing with a bcc ordering. A Z12 coordination is present around Cu, and higher coordinated Frank-Kasper polyhedra around Ti. The FFS and Frank-Kasper coordination is found to be the highest for the equiatomic Cu$_{50}$Ti$_{50}$ composition, which gives an explanation for the experimentally found viscosity maximum for this composition. Also, the general GFA of the binary CuTi system might be attributed to the complex local structures.

\section{Computational Background\label{sec:exp}}
The structural and dynamic characteristics of binary Cu$_{76}$Ti$_{24}$, Cu$_{50}$Ti$_{50}$, and Cu$_{31}$Ti$_{69}$ liquid alloys, within a temperature range of 1000 to 1700 K were investigated using AIMD simulations within the framework of the density functional theory (DFT), employing the Vienna \textit{ab initio} simulation package (VASP) \cite{Kresse_1993}. The electron-ion interaction was represented by the project augmented wave (PAW) potentials \cite{Blochl1994, Kresse1999}. For exchange and correlation effects, the generalized gradient approximation (GGA) in the Perdew, Burke, and Ernzerhof (PBE) formulation was applied \cite{Perdew1981}. The cutoff energy in plane-wave expansion was taken as the default one for Cu, namely 295 eV. For such large supercells in the liquid state, only the $\Gamma_0$ point was used for sampling the Brillouin zone. Phase space trajectories were generated through dynamical simulations by numerical integration of Newton's equations of motion using the standard Verlet algorithm in the velocity form, with a time step of 1.5 fs. All simulations were conducted within cubic boxes containing N = 256 atoms, with periodic boundary conditions applied in all spatial directions. The number of atoms for each species in the simulations is listed in Table \ref{tab:number_atoms}. The volume $V$ of each simulation was adjusted to match experimentally determined densities. The respective simulation box sizes are listed in Table \ref{tab:number_atoms} as well. Temperature control was achieved using a Nos\'e thermostat within the canonical ensemble (constant N, V, and T) \cite{Nose_1984}, with thermalization performed for a minimum of 20 ps, guided by the relaxation time determined from the self-intermediate scattering functions $F_S(q,t)$ \cite{Binder_2011}. Structural and dynamic properties were extracted from simulations spanning from 12 to 40 ps for each temperature. The shear viscosity was calculated from the diffusion coefficient using the Stokes-Einstein relation.
\begin{table}[tb]
\caption{\label{tab:number_atoms}%
Number of atoms of each species for all liquid Cu-Ti alloys used in the corresponding VASP simulations.
The total number of atoms is N = 256.}
\begin{ruledtabular}
\begin{tabular}{ccccccc}
&
\multicolumn{2}{c}{Cu$_{31}$Ti$_{69}$}&\multicolumn{2}{c}{\textrm{Cu$_{50}$Ti$_{50}$}}&\multicolumn{2}{c}{Cu$_{76}$Ti$_{24}$}\\
& Cu & Ti & Cu & Ti & Cu & Ti \\
\colrule
number of atoms & 79 & 177 & 128 & 128 & 195 & 61 \\

box size (1000 K) & \multicolumn{2}{c}{16.174 \AA} & \multicolumn{2}{c}{15.845 \AA} & \multicolumn{2}{c}{15.331 \AA}\\

box size (1100 K) & \multicolumn{2}{c}{16.211 \AA} & \multicolumn{2}{c}{15.888 \AA} & \multicolumn{2}{c}{15.390 \AA}\\

box size (1200 K) & \multicolumn{2}{c}{16.248 \AA} & \multicolumn{2}{c}{15.932 \AA} & \multicolumn{2}{c}{15.450 \AA}\\

box size (1300 K) & \multicolumn{2}{c}{16.287 \AA} & \multicolumn{2}{c}{15.985 \AA} & \multicolumn{2}{c}{15.510 \AA}\\

box size (1400 K) & \multicolumn{2}{c}{16.329 \AA} & \multicolumn{2}{c}{16.031 \AA} & \multicolumn{2}{c}{15.559 \AA}\\

box size (1500 K) & \multicolumn{2}{c}{16.363 \AA} & \multicolumn{2}{c}{16.067 \AA} & \multicolumn{2}{c}{15.634 \AA}\\

box size (1600 K) & \multicolumn{2}{c}{16.401 \AA} & \multicolumn{2}{c}{16.112 \AA} & \multicolumn{2}{c}{15.697 \AA}\\

box size (1700 K) & \multicolumn{2}{c}{16.441 \AA} & \multicolumn{2}{c}{16.159 \AA} & \multicolumn{2}{c}{15.761 \AA}\\

\end{tabular}
\end{ruledtabular}
\end{table}
Equilibrated trajectories at each temperature were sampled regularly to obtain ten configurations and relax then the inherent structure (IS) \cite{Stillinger_1982}. This step is crucial for decoupling vibrational motion from underlying structural properties by minimizing the potential-energy surface of the atoms, thereby reducing noise in the CNA index calculations. Structural analysis was conducted using the common neighbor analysis (CNA) method, implemented through various Python modules of the Open Visualization Tool (OVITO) \cite{Faken_1994, Stukowski_2010}.

\section{Results and Discussion}
\subsection{A. Structural Properties}
The reliability of the performed AIMD simulations is verified by comparison with already existing experimental data. Figure~1 shows the total structure factor $S(q)$ for the binary Cu$_{76}$Ti$_{24}$, Cu$_{50}$Ti$_{50}$, and Cu$_{31}$Ti$_{69}$ liquid alloys as well as for pure liquid Cu and Ti, obtained from experiments and AIMD. The experimental data were obtained using X-ray diffraction and electrostatic levitation \cite{Kreuzer_2024}. For $S(q)$ of the Cu-rich Cu$_{76}$Ti$_{24}$ composition only simulation results are available. The binary alloys were measured at 1273~K, the pure Cu and Ti at 1398~K and 1848~K, respectively. The AIMD were performed at similar temperatures (1300 K~for the binary alloys and 1398~K and 1848~K for pure Cu and Ti, respectively). Both, AIMD and experimental data agree very well regarding the position and amplitude of the first peak and the subsequent oscillations. Generally, this is a first indication of the reliability of the performed simulations.

\begin{figure}[tb]
\includegraphics[width=0.9\columnwidth]{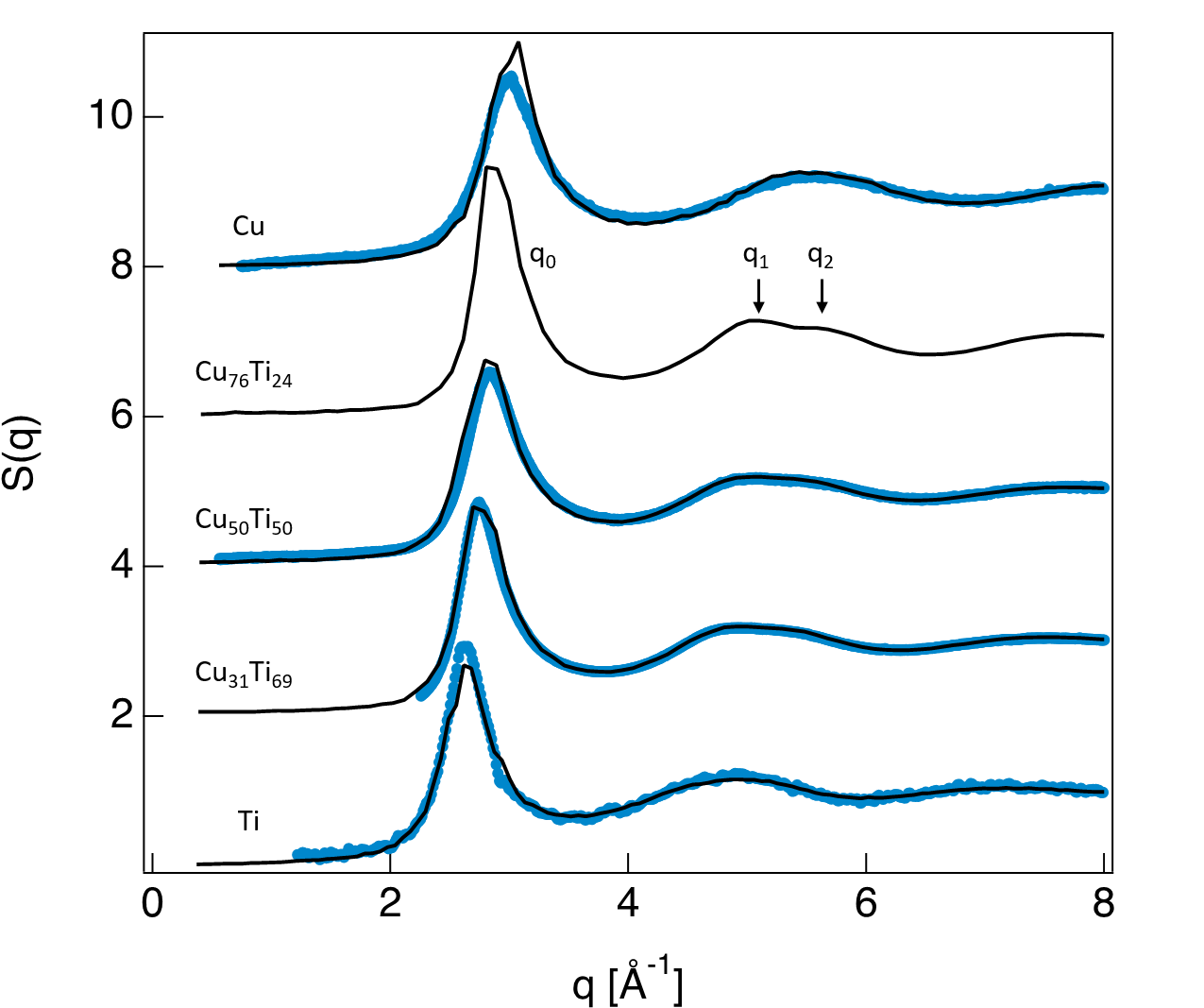}
\caption{AIMD calculations (black lines) of the total x-ray strucure factor S(q) in liquid Cu$_x$Ti$_{1-x}$ (from bottom to top x = 0, 0.31, 0.5, 0.76, 1). Experimental data (blue points) are shown for comparison. The binary alloys have been measured at 1273~K \cite{Kreuzer_2024}, while the pure Cu and Ti have been measured at 1398~K and 1848~K, respectively \cite{Holland-Moritz_2023, Holland-Moritz2007}. The AIMD have been performed at similar temperatures (1300 K for the binary alloys and 1398 K and 1848 K for pure Cu and Ti, respectively). \label{fig:all_compositions_S(q)}}
\end{figure}

For all three investigated Cu-Ti compositions, a shoulder at the second peak of the structure factor $S(q)$ is observed. Especially for Cu$_{76}$Ti$_{24}$, this shoulder is strongly pronounced. This feature in the scattering data suggests a preferred icosahedral SRO (ISRO) in the melt. Within the Ginzburg-Landau theory to describe the ISRO in undercooled liquids and metallic glasses \cite{Sachdev1984}, it can be further investigated by calculating the ratio of the position of the first and second peak maxima $q_0$ and $q_2$ as well as the respective shoulder $q_1$ (as is also shown in Figure~\ref{fig:all_compositions_S(q)}). For a perfect ISRO, values of $q_1/q_0 \approx$ 1.7 and $q_2/q_0 \approx$ 2 are expected, whereas the calculated values for the respective compositions are listed in Table \ref{tab:q_ratio} and are slightly larger (in the case of $q_1/q_0$) or lower ($q_2/q_0$). Thus, a non-negligible fraction of ISRO is present within all three investigated Cu-Ti compositions, however not a perfect one, as is expected for a binary alloy with atoms of different sizes.

\begin{table}[tb]
\caption{\label{tab:q_ratio}%
Ratio of $q_1/q_0$ and $q_2/q_0$ for all Cu-Ti compositions under investigation.}
\begin{ruledtabular}
\begin{tabular}{cccc}
& Cu$_{31}$Ti$_{69}$ & Cu$_{50}$Ti$_{50}$ & Cu$_{76}$Ti$_{24}$ \\
$q_1/q_0$ & 1.79 & 1.78 & 1.77 \\
$q_2/q_0$ & 1.91 & 1.84 & 1.90  
\end{tabular}
\end{ruledtabular}
\end{table}

\begin{figure}[tb]
\includegraphics[width=0.9\columnwidth]{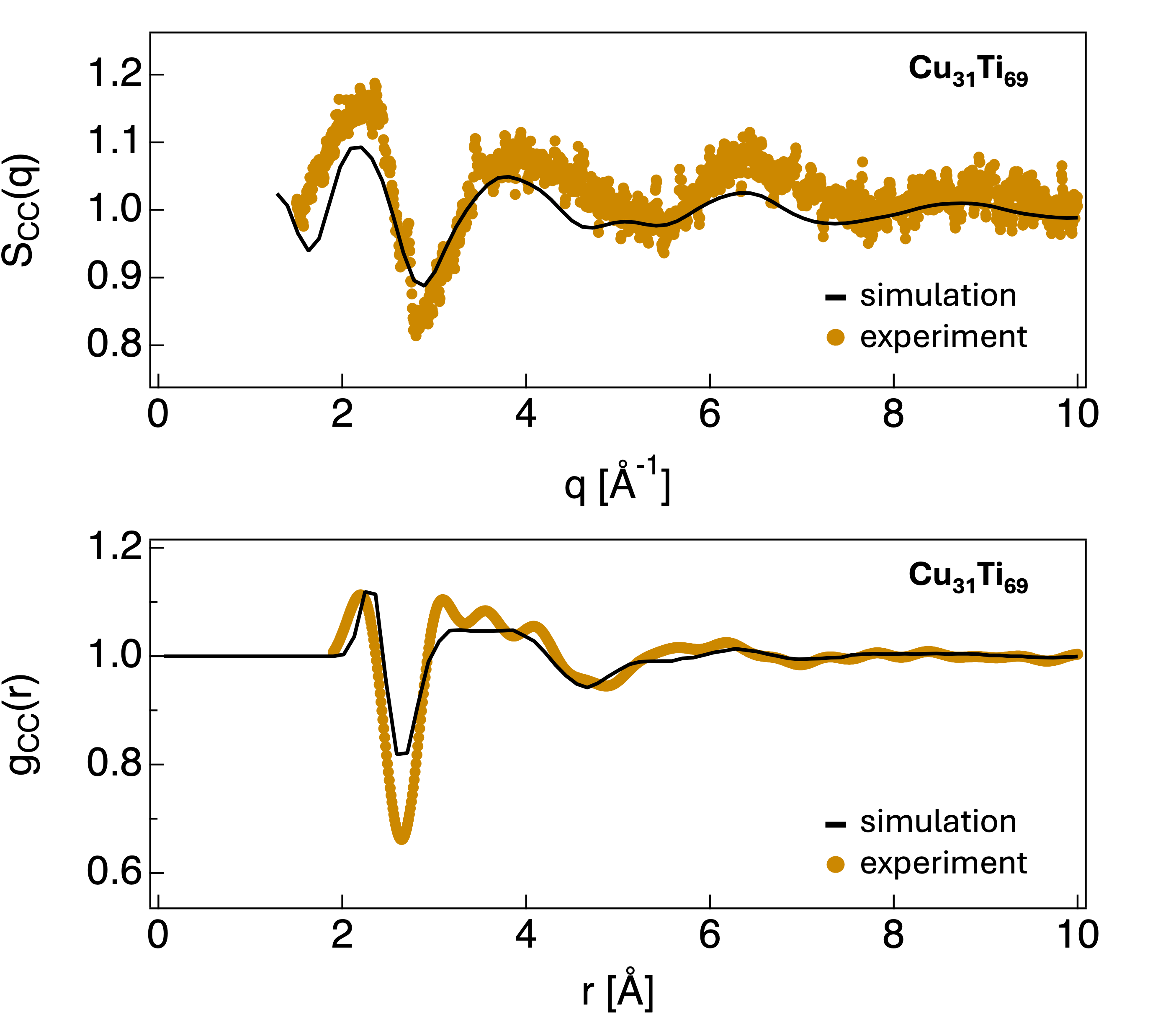}
\caption{Simulated (black lines) concentration-concentration structure factor S$_\text{CC}(q)$ (top) and pair-correlation function g$_\text{CC}(r)$ (bottom) of Cu$_{31}$Ti$_{69}$ at 1300~K. Experimental data (orange points) of $S_\text{CC}(q)$ and $g_\text{CC}(r)$ of Cu$_{31}$Ti$_{69}$ from neutron diffraction experiments (T = 1273~K) are shown for comparison \cite{Kreuzer_2024}.\label{fig:Cu31Ti69_Scc_gcc}}
\end{figure}

Furthermore, AIMD calculations of the partial concentration-concentration structure factor $S_\text{CC}(q)$ and pair-correlation function $g_\text{CC}(r)$ are compared with experimental results that have been obtained by neutron diffraction (again in combination with electrostatic levitation) in Figure~\ref{fig:Cu31Ti69_Scc_gcc}. Analogously to the total structure factor $S(q)$, the simulated  $S_\text{CC}(q)$  and $g_\text{CC}(r)$  feature a good agreement with the experimental data. This is remarkable, since $S_\text{CC}(q)$ reflects on the chemical ordering, and thus, its Fourier transformation gives an idea of the distribution of the chemical species. The good overlap of experimental data and simulation demonstrates that the used AIMD frameworks not only can reproduce the overall structural properties but also correctly considers present chemical interactions within the Cu-Ti system. 

In the following, the structural properties i.e., local structures, coordination numbers, and symmetries of Cu-Ti melts from AIMD are discussed. Subsequently, the melt dynamics through the mean self-diffusion coefficient and shear viscosity, as well as the correlation of the dynamic properties with the local melt structure will be analyzed.

\begin{figure*}[tb]
\includegraphics[width=1.8\columnwidth]{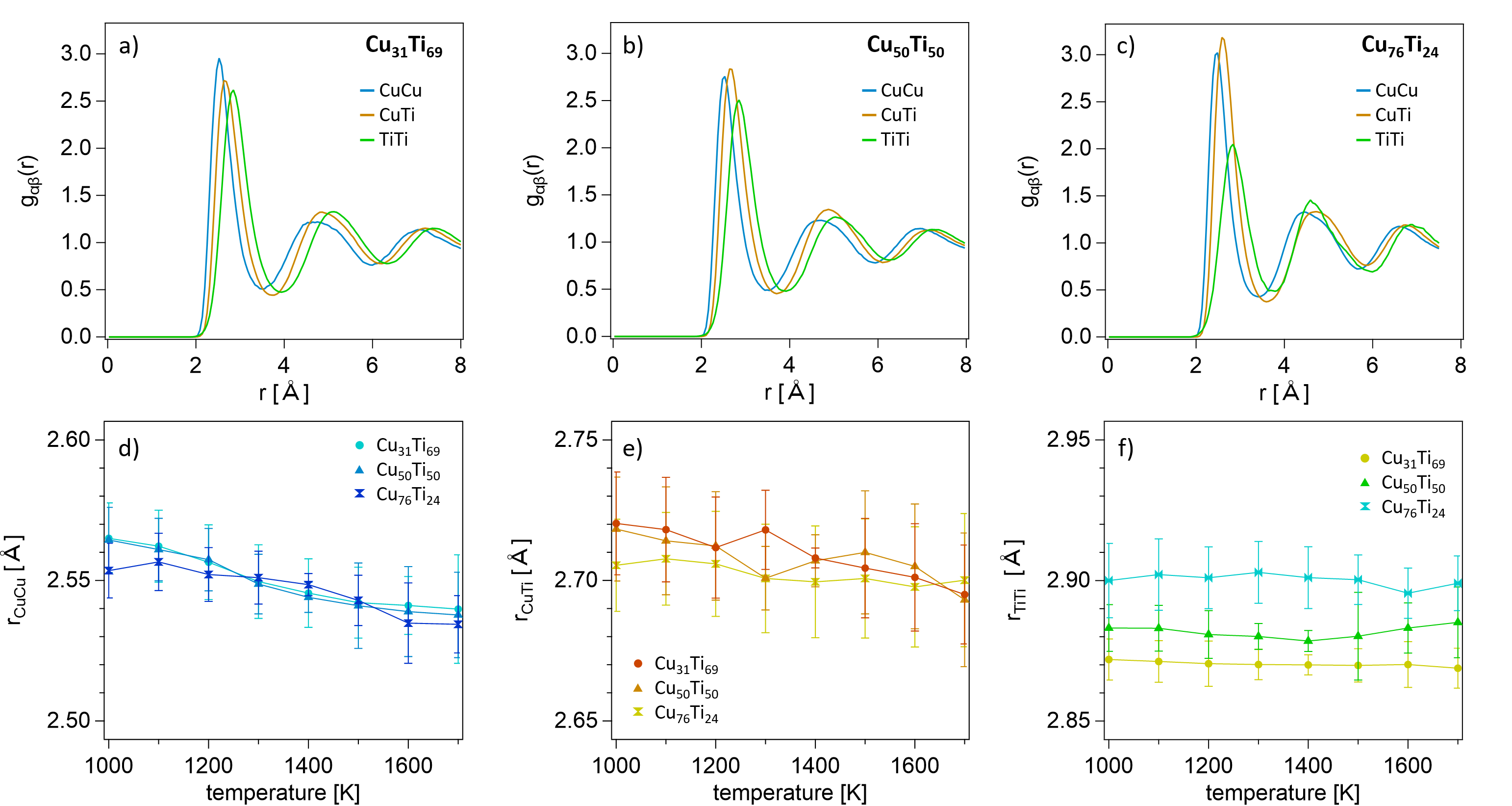}
\caption{Partial pair-correlation functions of (a) Cu$_{31}$Ti$_{69}$, (b) Cu$_{50}$Ti$_{50}$, and (c) Cu$_{76}$Ti$_{24}$ at a temperature of 1300 K. In the bottom row, the positions of the first maxima of $r_{\alpha\beta}$, which represents the interatomic particle distance between (d) CuCu, (e) CuTi, and (f) TiTi, respectively, is shown as a function of temperature. \label{fig:CuTi_rdf}}
\end{figure*}

Partial pair-correlation functions $g_{\alpha\beta}(r)$ obtained from the AIMD simulations are shown in Figure~\ref{fig:CuTi_rdf}(a)-(c) at an exemplary temperature of 1300~K for the three investigated compositions. Figure~\ref{fig:CuTi_rdf}(d)-(f) shows the first peak maxima position $r_{\alpha\beta}(T)$ of the respective $g_{\alpha\beta}(r)$ as a function of temperature are plotted. $r_{\alpha\beta}(T)$ represents the average first neighbor distance and thus, the Cu-Cu, Cu-Ti, and Ti-Ti bond lengths, respectively. For all three compositions, we find the Cu-Cu distances to be shortest and Ti-Ti distances to be longest, which is due to the larger atomic radius of Ti ($r_\text{Ti} = 1.324$~\AA~and $r_\text{Cu} = 1.173$~\AA). While the Cu-Cu and Cu-Ti distances feature no composition dependence, the average Ti-Ti distance (Figure~\ref{fig:CuTi_rdf}f) displays increasing distances upon increasing Cu concentration. The lowest and highest Ti-Ti distance is observed for the Ti-rich and Ti-poor Cu$_{31}$Ti$_{69}$ ($r_\text{TiTi} \approx$ 2.87~\AA) and Cu$_{76}$Ti$_{24}$ ($r_\text{TiTi} \approx$ 2.90~\AA) composition, respectively. This observation can be attributed to extraordinary high coordination numbers around Ti, in particular for the Cu$_{76}$Ti$_{24}$ composition, as will be explained later.
Figure~\ref{fig:CuTi_rdf}(d)-(e) highlights that with increasing temperature, the average distances display a small decrease, which is slightly more pronounced for Cu-Cu and Cu-Ti. Such a trend was also observed for pure elements and can be correlated with an increase in coordination number, as will be discussed later in the text \cite{Jakse_2008b}. The values for $r_\text{TiTi}$ remain constant with temperature. To put these values into context, the atomic distances between Cu and Ti in the pure melts are slightly smaller as the respective distances in the alloys ($r_\text{CuCu, pure}$ = 2.48~\AA~and $r_\text{TiTi, pure}$ = 2.80~\AA) \cite{Schenk_2002, Holland-Moritz2007, Holland-Moritz_2023}. The amplitudes of $g_{\alpha\beta}(r)$ relatively to each other change, showing an evolution from an ideal mixing at low Cu content to a more pronounced hetero-coordination with increasing Cu composition as revealed by the amplitude of $g_\text{CuTi}$ which increases, whereas those of $g_\text{TiTi}$ decreases. As a matter of fact, a monotonous decrease from the amplitude of $g_\text{CuCu}(r)$ to $g_\text{CuTi}(r)$ and $g_\text{TiTi}(r)$ ($A_{g_\text{CuCu}} > A_{g_\text{CuTi}} > A_{g_\text{TiTi}}$) is observed for Cu$_{31}$Ti$_{69}$ (see Figure \ref{fig:CuTi_rdf}a), corresponds typically to an ideal solution behavior \cite{Hansen2006}. The more pronounced $g_\text{CuTi}(r)$ for the other two compositions Cu$_{50}$Ti$_{50}$ and Cu$_{76}$Ti$_{24}$, suggests a deviation from such an ideal solution behavior. In the literature, the solution behavior was studied experimentally and different results were obtained: Amore and coworkers observed a strong deviation from the ideal solution behavior, with a maximum at the Ti-rich side \cite{Amore2011b, Amore2013}. In contrast, some of us recently reported a solution behavior that is closer to the ideal case, with a maximal deviation found for intermediate Ti contents. One should note that no data for a Cu-rich alloy was shown there \cite{Kreuzer_2024}. The different results on the solution behavior reported in references \cite{Amore2013} and \cite{Kreuzer_2024} have been referred to different measurement and sample processing techniques, which also highlight the sensitivity and complexity of the Cu-Ti system. The results obtained from AIMD simulation feature a better agreement with the experimental results from reference \cite{Kreuzer_2024}. 

As already mentioned, with decreasing Ti-content, the Cu-Ti affinity becomes more dominant. This could be an explanation for the experimentally obtained trend of the melt viscosity that features a maximum for Cu$_{50}$Ti$_{50}$ as will be seen below. There, the high affinity of Cu-Ti and a higher number of Cu-Ti pairs as compared to Cu$_{76}$Ti$_{24}$ apparently lead to a significant slow-down of the viscosity as was experimentally observed \cite{Kreuzer_2024}.

\subsection{Coordination numbers}
\begin{figure*}[tb]
\includegraphics[width=1.8\columnwidth]{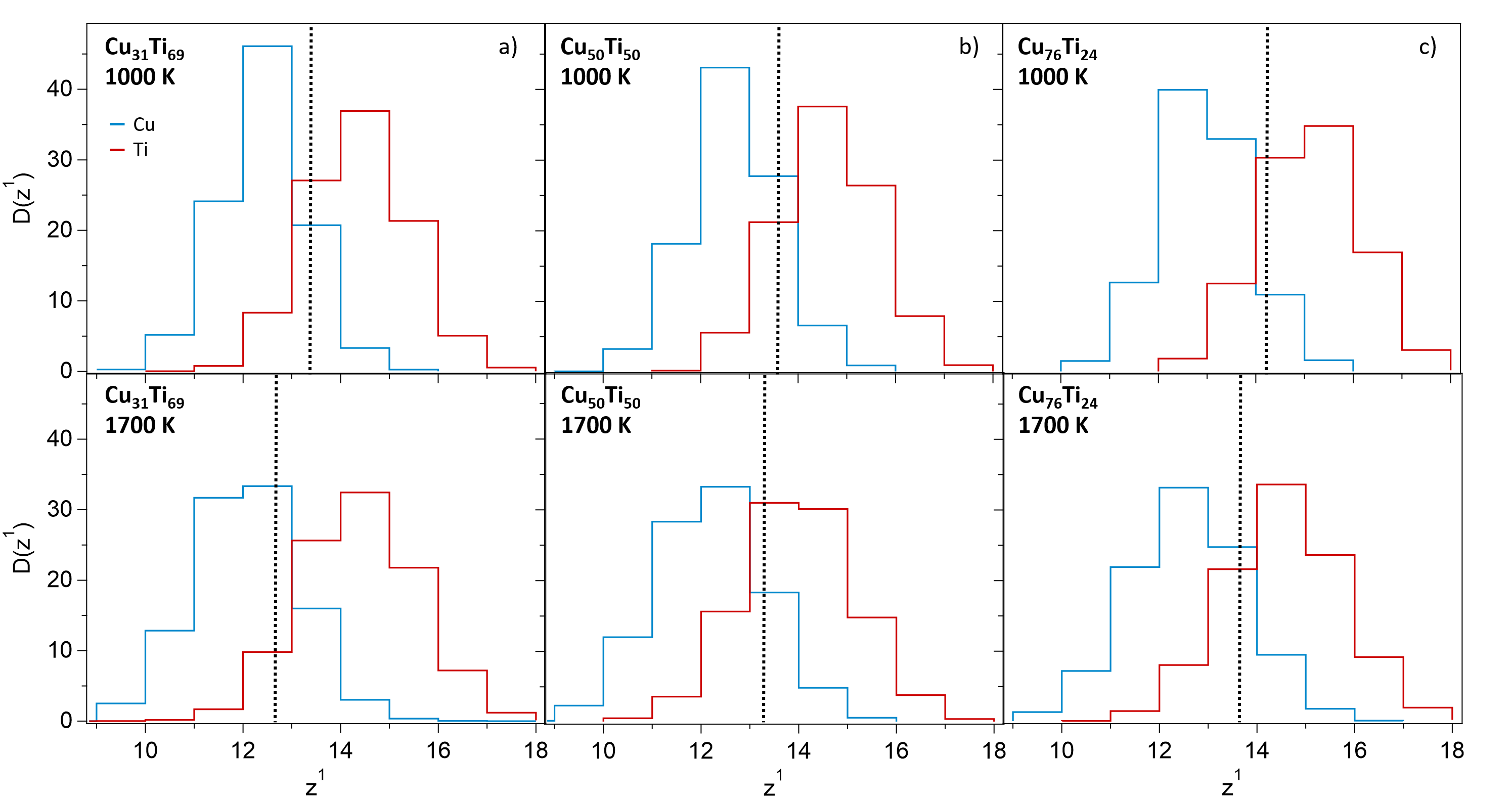}
\caption{Distribution of the coordination numbers $z^1$ for typical configurations of (a) Cu$_{31}$Ti$_{69}$, (b) Cu$_{50}$Ti$_{50}$, and (c) Cu$_{76}$Ti$_{24}$ in the deeply undercooled (1000 K, top row) and liquid state (1700 K, bottom row) for Cu (blue) and Ti (red). The vertical black dashed lines indicate the average coordination number for the respective Cu-Ti composition. \label{fig:CuTi_Dz}}
\end{figure*}

The coordination number distributions are determined by counting the number of neighboring Cu and Ti atoms within the first coordination shell defined by the first minimum of the corresponding partial pair-correlation functions $g_{\alpha\beta}(r)$. They are plotted for both Cu and Ti in the liquid (1700 K) and undercooled region (1000 K) in Figure \ref{fig:CuTi_Dz}. They give a first indication of the nature of the SRO. Upon cooling, we observe an increase in the coordination numbers for both Cu and Ti. This is in contrast to e.g., the Cu-Zr system, where the opposite behavior has been found \cite{Jakse_2008a}. In both, the liquid and undercooled region, Cu features mainly a Z12 coordination, i.e., a large fraction of the Cu atoms are surrounded by a simple icosahedron as will become clearer from the CNA discussion below. This also agrees with the calculated ratios of the respective peak positions of the total structure factor $S(q)$ listed in Table \ref{tab:q_ratio}, which already indicated the presence of an ISRO. An ISRO is not compatible with a long-range periodicity of the crystalline phase and might already explain the large degree of undercooling of the Cu-Ti system and thus, the potentially good GFA of Cu-Ti-based alloys \cite{Frank_1952}.

\subsection{Common neighbor analysis}

By performing a common neighbor analysis (CNA) on inherent configurations of the melt, an even more detailed image of the topology of the polyhedra surrounding the Cu and Ti atoms is obtained. 

The most important bonded pairs for all compositions in the undercooled and liquid state (1000 and 1700 K) are reported in Table \ref{tab:CNA}. The numbers within the square brackets represent specific structural characteristics of atomic clusters. After identifying all neighboring atoms with a specified cutoff radius $R_c$ corresponding to the first minimum of the partial pair-correlation function around a selected atom, each pair formed by the atom and its neighbors is analyzed. The goal is to characterize the common neighbors of each pair in the first coordination shell using a set of three indices $jkl$: $j$ represents the number of neighbors shared by both atoms, $k$ denotes the number of bonds between these shared neighbors (a bond is considered present if the distance between neighbors is smaller than $R_c$), and $l$ indicates the number of bonds in the longest chain among the common neighbors. Before discussing the found cluster symmetries, it is worth highlighting the absence or negligible amounts of [421] and [422] symmetries, which would signal the presence of a fcc and hcp ordering like in solid Cu and Ti, respectively. Thus, it is assured that a fully liquid system is present during our simulations. In contrast, an underlying bcc structure, which is indicated by the small, but non-negligible amount of [444] symmetry, cannot be fully excluded, when associated to [666] pairs.

In general, upon undercooling, an increase for all pairs, except for the [433] one, is observed, indicating an increasing order of the melt. This agrees with the coordination number, which also, on average, increase for both Cu and Ti upon cooling. In general, a FFS i.e., an ISRO, which is represented by the [555] pairs and its distorted version [554], is already present in the liquid state for both Cu and Ti, while upon undercooling, a strong increase, especially for Cu is observed. With the high fraction of Z12 coordination for Cu, the strong occurrence of the [555] symmetry, especially in the undercooled melt is also a strong indicator for the good undercooling rate and potentially high GFA of the Cu-Ti and its multi-component systems.

Associated to [555], the CNA reveals a significant amount of [666] symmetry, essentially around Ti. This refers to high coordination polyhedra, most likely Frank-Kasper polyhedra. Upon cooling, this [666] symmetry increases especially for Ti, which is in good agreement with the high coordination numbers found for Ti. In addition, a smaller, but non-negligible amount of [444] is found highlighting a competition between the Frank-Kasper and bcc ordering. The only type of symmetry that is decreasing upon cooling is the [433] ordering, which can be seen as distorted pair with a five- or four-fold symmetry, thus indicating that the liquid becomes more structured. For all compositions, the five-fold [555] symmetry increases the strongest upon decreasing temperature, and thus, we can conclude that the [433] ordering in the melt transforms mainly into a [555] symmetry around Cu, while a significantly smaller fraction transforms into a [444] ordering, typical to the bcc ordering. Overall, Cu$_{50}$Ti$_{50}$ features the highest amount of a five-fold and [666] symmetry. In addition, upon undercooling the melt, the amount of [433] pairs decreases the strongest for Cu$_{50}$Ti$_{50}$. Consequently, the FFS symmetry of Cu$_{50}$Ti$_{50}$ increases the strongest upon cooling. The detailed structural picture can be used to explain the melt dynamics, which are the slowest for Cu$_{50}$Ti$_{50}$. While an ISRO is known to slow down melt dynamics \cite{Schenk_2002}, the complex coordination scenario, especially around Ti contributes as well: Atom diffusion would affect many atoms coordinated in Frank-Kasper polyhedra around the center atom (Z15 and Z16) and is therefore, not favored, which in turn slows down the melt dynamics.

In the undercooled melt, with decreasing Ti content, the fraction of Z12 coordination slightly decreases, whereas a Z13 coordination, meaning a distorted five-fold ISRO becomes more dominant for Cu. Over the entire temperature range, the coordination numbers for Ti are significantly higher than for Cu, suggesting a more complex coordination polyhedron than the simple icosahedron. We observe a non-negligible amount of Z15 and Z16 coordination for Ti, which can be addressed to Frank-Kasper polyhedra due to the presence of [666] pairs. For Cu we also observe higher coordination numbers (Z13 and Z14), however, to a significant lower degree. Thus, the complexity of the coordination scenario around Ti is apparently higher than for Cu. This applies in particular for Cu$_{76}$Ti$_{24}$, where the fractions of Z15 and Z16 coordination is the highest. By taking into account the strong amplitude of the first peak of $g_{CuTi}$, which indicates the preferential presence of heterogeneous Cu-Ti pairs, the high coordination numbers around Ti also can explain the increasing Ti-Ti distances with decreasing Ti-content (cf. Figure \ref{fig:CuTi_rdf}f). With more Ti atoms that are preferably surrounded by Cu atoms, the average Ti-Ti distance increases. The weak amplitude of the first maxima of $g_{TiTi}$ especially for Cu$_{76}$Ti$_{24}$ confirms this assumption. 

In summary, CNA allows us to extract a detailed three-dimensional picture about the short-range order and its evolution upon decreasing temperature. Generally, the way in which the SRO evolves upon undercooling is similar for all investigated Cu-Ti alloys. The FFS is found to be the most important feature: For Cu, the FFS increases linearly with increasing Cu content, while around Ti the FFS is the highest at intermediate Ti contents. In total, the FFS fraction is highest for Cu$_{50}$Ti$_{50}$, which provides an explanation for the slowest melt dynamics, which was found for this composition. High coordination Frank-Kasper polyhedra are present in particular around Ti, and an underlying bcc ordering is obtained as well suggesting a competition between the FFS and the bcc orderings \cite{Jakse2003}. Interestingly, the [666] fraction is the highest for an intermediate Ti content, and lead to the assumption that these Frank-Kasper polyhedra also contribute to the slow-down of the Cu-Ti system.

\begin{table*}[tb]
\caption{\label{tab:CNA}%
Common-neighbor analysis for Cu$_{31}$Ti$_{69}$, Cu$_{50}$Ti$_{50}$, and Cu$_{76}$Ti$_{24}$ at T = 1000 and 1700 K (undercooled and stable liquid region). The error bars for the calculated pair abundances are of the order of 0.01. Please note that the given values for [666] also include the fraction of the distorted six-fold symmetry [677].}
\begin{ruledtabular}
\begin{tabular}{cccccccccccccc}
&
\multicolumn{4}{c}{Cu$_{31}$Ti$_{69}$}&\multicolumn{4}{c}{\textrm{Cu$_{50}$Ti$_{50}$}}&\multicolumn{4}{c}{Cu$_{76}$Ti$_{24}$}\\
 & \multicolumn{2}{c}{Cu} & \multicolumn{2}{c}{Ti} & \multicolumn{2}{c}{Cu} & \multicolumn{2}{c}{Ti} & \multicolumn{2}{c}{Cu} & \multicolumn{2}{c}{Ti} \\
\textrm{CNA (\%)} & 1000 K & 1700 K & 1000 K & 1700 K & 1000 K & 1700 K & 1000 K & 1700 K & 1000 K & 1700 K & 1000 K & 1700 K\\
\colrule
[421] & 0.00 & 1.72 & 0.00 & 1.44 & 1.14 & 2.23 & 0.00 & 1.97 & 1.18 & 2.79 & 0.00 & 1.89 \\ \relax
[422] & 1.86 & 4.53 & 1.89 & 3.57 & 2.30 & 4.35 & 1.71 & 3.26 & 1.95 & 5.31 & 1.71 & 4.43\\ \relax
[433] & 14.4 & 18.2 & 11.7 & 14.8 & 14.4 & 19.0 & 9.79 & 15.2 & 14.7 & 19.2 & 10.6 & 15.9 \\ \relax
[444] & 9.34 & 8.91 & 7.36 & 7.19 & 8.72 & 7.71 & 7.16 & 7.00 & 9.70 & 7.23 & 8.28 & 6.27 \\ \relax
[544] & 12.6 & 14.4 & 14.8 & 15.2 & 13.8 & 13.7 & 14.7 & 14.8 & 12.6 & 14.7 & 14.0 & 16.7 \\ \relax
[555] & 40.4 & 19.1 & 26.5 & 21.3 & 37.2 & 18.1 & 35.0 & 19.0 & 34.2 & 17.5 & 31.2 & 19.1 \\ \relax
[666] & 11.3 & 9.72 & 18.1 & 13.9 & 12.3 & 9.76 & 19.9 & 14.2 & 12.7 & 9.24 & 18.2 & 13.8 
\end{tabular}
\end{ruledtabular}
\end{table*}

\subsection{Dynamic properties of Cu-Ti alloys}

\begin{figure}[tb]
\includegraphics[width=0.9\columnwidth]{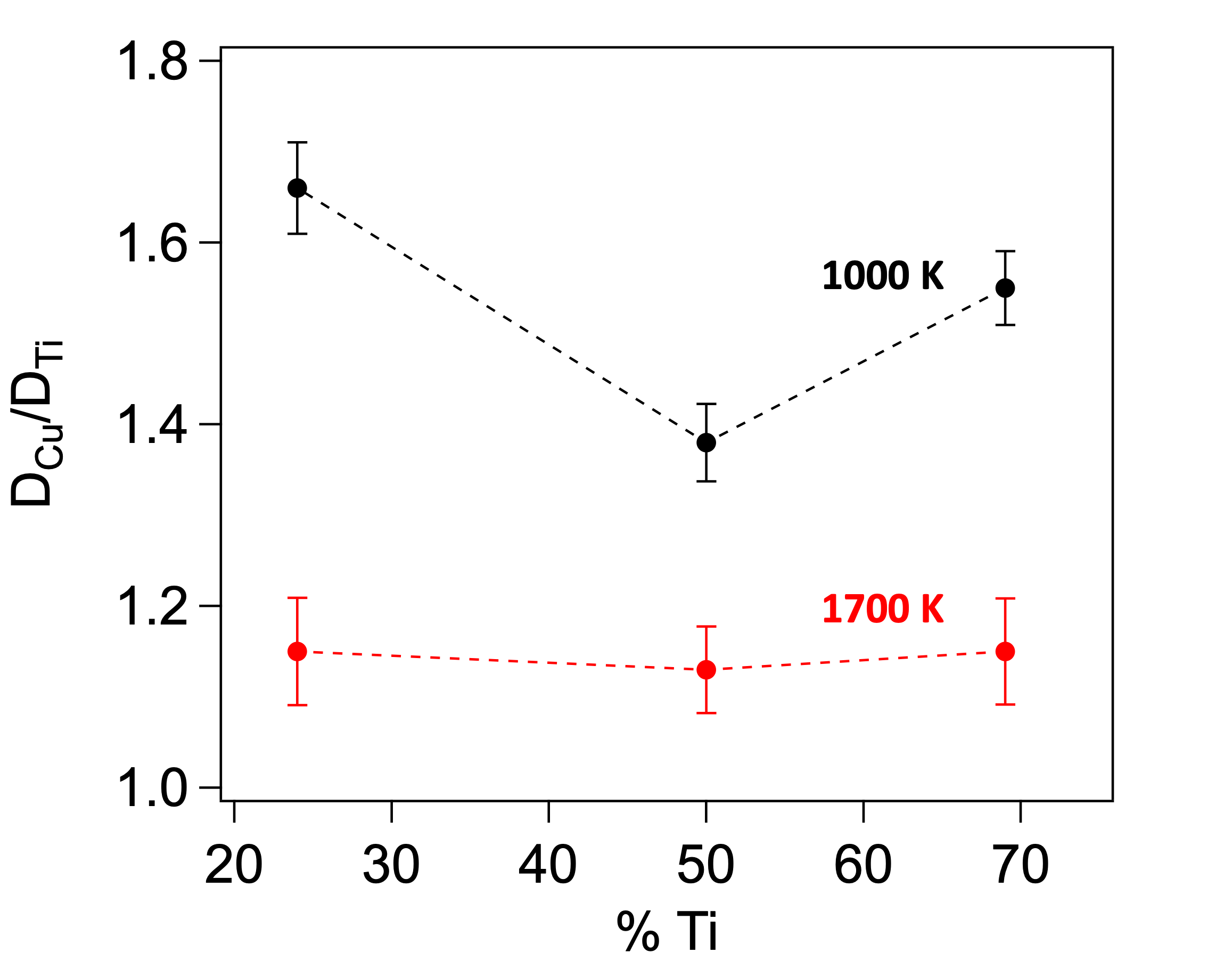}
\caption{Ratio $D_{Cu}$/$D_{Ti}$ from AIMD simulations depending on the composition at 1000 K (black) and 1700 K (red). The dashed lines are guides to the eye. \label{fig:D_ratio}}
\end{figure}

Figure \ref{fig:D_ratio} shows the ratio of Cu and Ti diffusion coefficients $D_{Cu}$/$D_{Ti}$ as a function of Ti content, calculated with AIMD simulations at 1000 and 1700 K. Thereby, the dynamical decoupling of the atomic species and its dependence on the composition can be addressed, which is not possible experimentally. At high temperatures (1700 K), the $D_{Cu}$/$D_{Ti}$ ratio remains constant around a value of 1.15 over the entire composition range. The situation changes when decreasing the temperature. At 1000 K significantly higher values of the $D_{Cu}$/$D_{Ti}$ ratio for all Cu-Ti compositions are obtained, suggesting an overall decoupling of the diffusion coefficients. In particular, the high values at the Ti-rich side, indicate weak Cu-Ti interactions: the smaller Cu atoms are weakly bound to the larger Ti atoms, and can move more readily through the melt, thereby leading to the decoupling of the diffusion coefficients. This is in contrast to other systems, e.g., binary Ni-Zr, where stronger interactions between Ni and Zr, and no decoupling at the Ni-rich side have been found \cite{Nowak2017a, Basuki2017}. Here, with increasing Ti content, the $D_{Cu}$/$D_{Ti}$ ratio first decreases from 1.66 $\pm$ 0.04 to 1.38 $\pm$ 0.04 for Cu$_{50}$Ti$_{50}$, before increasing to 1.55 $\pm$ 0.05 for Cu$_{31}$Ti$_{69}$. This observed composition-dependency results from the changing numbers of Cu-Ti pairs, which is the highest for the equiatomic Cu$_{50}$Ti$_{50}$ composition, where also the strongest $D_{Cu}$/$D_{Ti}$ is found. As a consequence of this, a large fraction of FFS and complex Frank-Kasper polyhedra forms. With other words, the melt dynamics are governed by the local structure, which is a consequence of the present chemical interactions, which shows a pronounced composition-dependence at a low temperature of 1000 K.

\begin{figure}[tb]
\includegraphics[width=0.9\columnwidth]{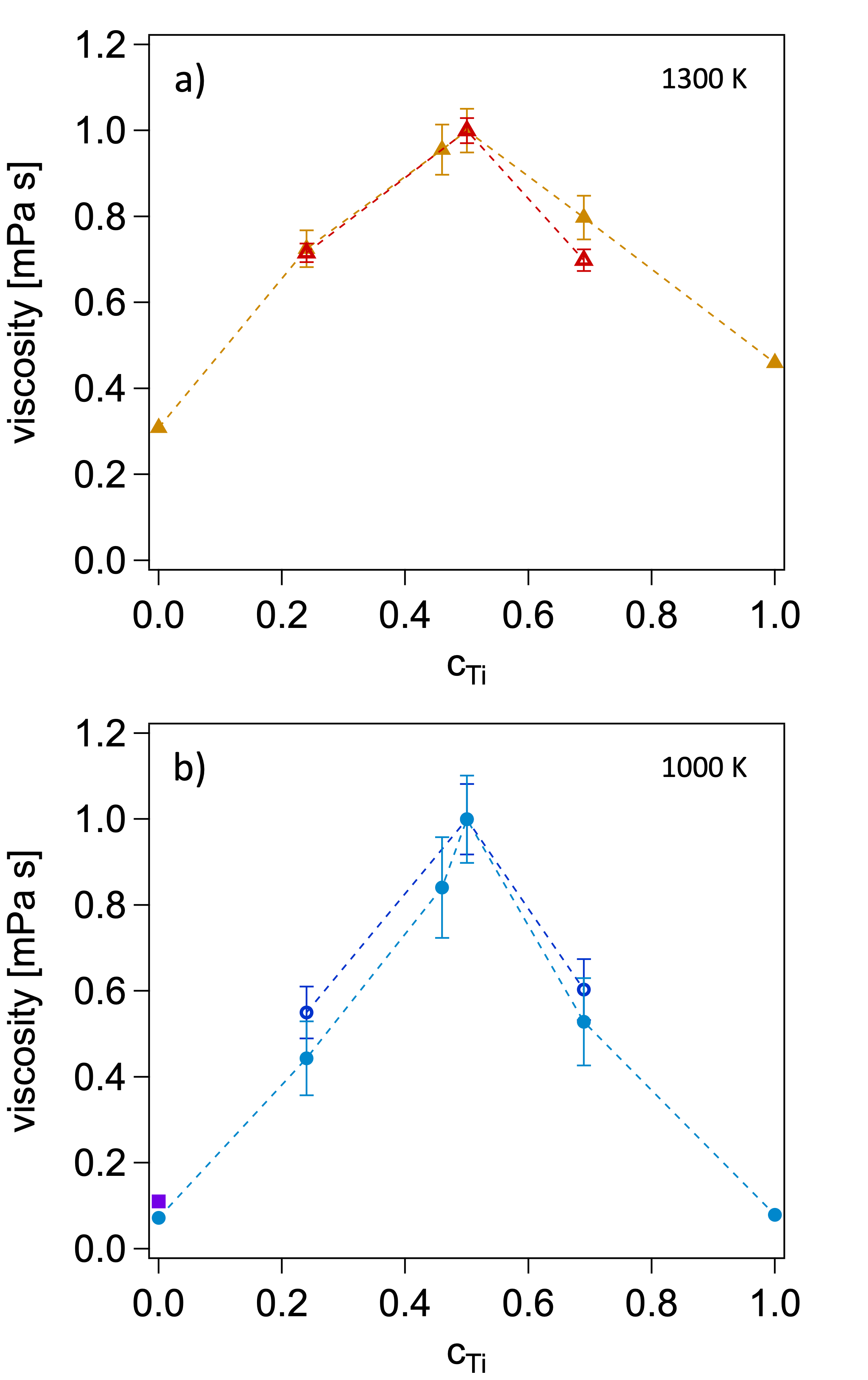}
\caption{Evolution of the calculated viscosity in the (a) liquid and (b) undercooled region at 1300 K (open orange triangles) and 1000 K (open blue circles), respectively, as a function of the Ti content, including error bars of 15 \%. The values have been normalized to the respective highest viscosity. Experimental data are shown as full symbols according to reference \cite*{Kreuzer_2024}, whereas the temperature slightly differs from those of the calculated viscosities (T = 1030 and 1273 K). The data for pure Cu and liquid has been extrapolated from experimental data that was available within an temperature range of 1360 to 1960 K for Cu and 1730 to 2120 for Ti \cite{Assael2010, Paradis2002, Ishikawa2012}. The viscosity of pure Cu at a temperature of approximately 1380 K, computed by a direct method using the transverse current-current correlation function $C_T(q,t)$ is shown as full purple square and was taken from reference \cite*{Jakse_2016}. \label{fig:viscosity}}
\end{figure}

In addition to the diffusion coefficients, the effect of the local structure and the present symmetry on the dynamic properties of stable and undercooled liquids can be investigated by calculating the respective viscosities. Figure \ref{fig:viscosity} shows the calculated viscosity as a function of Ti content for the stable and undercooled liquid at 1300 K and 1000 K, respectively. The viscosity has been calculated according to the Stokes-Einstein (SE) relation $\eta_E = k_BT/2\pi RD$, where $k_B$ is Boltzmann's constant, and $R$ the average interatomic particle distance at temperature $T$. Furthermore, $D$ stands for the mean self-diffusion coefficient. The calculated values are compared to the experimentally obtained viscosities, which are also plotted in Figure \ref{fig:viscosity} (closed symbols). Please note that the respective temperatures amount to 1273 K and 1030 K, and thus, differ slightly from the temperatures of the calculated values, which are 1300 K and 1000 K.

In general, the calculated values using the SE relation are lower than the experimentally measured ones. This is referred to the break-down of the SE relation: Upon alloying pure Cu and Ti, local structures evolve around the respective species as has been shown above. The familiar SE relation, which uses the effective radii from partial pair-correlation functions, is not capable of taking these local structures into account, and instead assumes a uniform diffusion of both Cu and Ti atoms. Thus, the SE relation is oversimplified and not valid in general for binary or multi-component alloys. However, a constant error is obtained with respect to the composition, which means that the obtained compositional trend is qualitatively correct. To allow for better comparison between the experimental and calculated values, we normalized all values to the highest viscosity. At both temperatures, 1000 and 1300 K, the AIMD simulations are able to reproduce the viscosity maximum at intermediate Ti contents. This further indicates a correct representation of the Cu-Ti correlation, as well as the local structures and coordination scenarios around Cu and Ti. In addition, the normalized viscosity of pure Cu at a temperature of approximately 1380 K is plotted in Figure \ref{fig:viscosity}b) (full purple square). This value was calculated by a direct method using the transverse current-current correlation function $C_T(q,t)$ as is described in references \cite*{Jakse_2013, Jakse_2016}. It has been shown that this approach leads to reliable values of the shear viscosity for metallic liquids with an error bar on the order of 0.2 mPa s. In the case of pure Cu, the computed value ($\approx$ 4 mPa s) is reasonably close to the experimentally found value ($\approx$ 6 mPa s), especially when considering that the calculated value has been derived at a higher temperature of 1380 K as compared to 1273 K for the experimental viscosity. Especially in the undercooled regime at 1000 K the viscosity of the equiatomic Cu$_{50}$Ti$_{50}$ composition was found to be the highest, which can be explained by the highest FFS fraction and complex coordination scenario, reflected by the high fraction of Frank-Kasper polyhedra. From AIMD the Cu$_{50}$Ti$_{50}$ was found to exhibit the highest fraction of FFS (around Cu and Ti) and thus might be responsible for the increased melt viscosity. Furthermore, also a maximum of the high coordination Frank-Kasper polyhedra was found for Cu$_{50}$Ti$_{50}$, which further promotes an increased viscosity. This is also consistent with the strongest Cu-Ti correlation of Cu$_{50}$Ti$_{50}$ found via the $D_{Cu}$/$D_{Ti}$ ratio (cf. Figure \ref{fig:D_ratio}.

\section{Conclusion}
First-principle based \textit{ab-initio} molecular-dynamics (AIMD) simulations have been performed for binary Cu$_x$Ti$_{1-x}$ (x = 0.31, 0.50, and 0.76) alloys over a wide temperature range, covering both liquid and undercooled state, to investigate the relationship of the local structure and dynamical properties. The common neighbor analysis (CNA) reveals increasing local order with decreasing temperature, dominated by a five-fold symmetry (FFS) around Cu. Especially in the undercooled region, higher coordination numbers that are compatible with Frank-Kasper polyhedra were obtained for Ti and to a lower degree also for Cu. In addition, the FFS and Frank-Kasper polyhedra compete with a bcc ordering. 
A maximum of FFS is found for the equiatomic Cu$_{50}$Ti$_{50}$ composition, while also the fraction of Frank-Kasper polyhedra around Ti is the highest. These local structures lead to strong Cu-Ti correlations and a maximum in the melt viscosity, which was found experimentally and via AIMD simulations. This correlation suggests that the melt viscosity is mainly slowed down by the FFS symmetry and the complexity of the local structure and coordination scenario, whereas attractive chemical interactions between Cu and Ti i.e., the preferred formation of Cu-Ti pairs, seem to play a subordinate role. Instead, a behavior close to an ideal-solution behavior was found, which agrees with recent experimental studies.
With the use of \textit{ab initio} simulations, the focus of this study was on local ordering and dynamical effects. In order to study long-range effects as well, a new approach could be used, which includes the development of Machine Learning (ML) interatomic potentials that are trained on appropriate sets of AIMD configurations. This allows to perform million-atom molecular dynamics (MD) simulations with the accuracy close to \textit{ab initio} \cite{Behler_2021, Jakse2023}.

\begin{acknowledgments}
We want to thank Prof. Dr. Winfried Petry for constructive discussions of the experimental Cu-Ti data, and Dr. Alaa Fahs for technical help with the AIMD simulations. Financial support provided by the Deutsche Forschungsgemeinschaft (DFG) via Grant No. 315677471 is gratefully acknowledged. Furthermore, we acknowledge the CINES, IDRIS and TGCC under Project No.  INP2227/72914/gen5054, as well as CIMENT/GRICAD for computational resources.  We acknowledge financial support under the French-German project PRCI ANR-DFG SOLIMAT (ANR-22-CE92-0079-01).  This work has been partially supported by MIAI@Grenoble Alpes (ANR-19-P3IA-0003). Discussions within the French collaborative network in artificial intelligence in materials science GDR CNRS 2123 (IAMAT) are also acknowledged.
\end{acknowledgments} 
\bibliographystyle{apsrev4-2}

\begin{thebibliography}{51}%
	\makeatletter
	\providecommand \@ifxundefined [1]{%
		\@ifx{#1\undefined}
	}%
	\providecommand \@ifnum [1]{%
		\ifnum #1\expandafter \@firstoftwo
		\else \expandafter \@secondoftwo
		\fi
	}%
	\providecommand \@ifx [1]{%
		\ifx #1\expandafter \@firstoftwo
		\else \expandafter \@secondoftwo
		\fi
	}%
	\providecommand \natexlab [1]{#1}%
	\providecommand \enquote  [1]{``#1''}%
	\providecommand \bibnamefont  [1]{#1}%
	\providecommand \bibfnamefont [1]{#1}%
	\providecommand \citenamefont [1]{#1}%
	\providecommand \href@noop [0]{\@secondoftwo}%
	\providecommand \href [0]{\begingroup \@sanitize@url \@href}%
	\providecommand \@href[1]{\@@startlink{#1}\@@href}%
	\providecommand \@@href[1]{\endgroup#1\@@endlink}%
	\providecommand \@sanitize@url [0]{\catcode `\\12\catcode `\$12\catcode
		`\&12\catcode `\#12\catcode `\^12\catcode `\_12\catcode `\%12\relax}%
	\providecommand \@@startlink[1]{}%
	\providecommand \@@endlink[0]{}%
	\providecommand \url  [0]{\begingroup\@sanitize@url \@url }%
	\providecommand \@url [1]{\endgroup\@href {#1}{\urlprefix }}%
	\providecommand \urlprefix  [0]{URL }%
	\providecommand \Eprint [0]{\href }%
	\providecommand \doibase [0]{https://doi.org/}%
	\providecommand \selectlanguage [0]{\@gobble}%
	\providecommand \bibinfo  [0]{\@secondoftwo}%
	\providecommand \bibfield  [0]{\@secondoftwo}%
	\providecommand \translation [1]{[#1]}%
	\providecommand \BibitemOpen [0]{}%
	\providecommand \bibitemStop [0]{}%
	\providecommand \bibitemNoStop [0]{.\EOS\space}%
	\providecommand \EOS [0]{\spacefactor3000\relax}%
	\providecommand \BibitemShut  [1]{\csname bibitem#1\endcsname}%
	\let\auto@bib@innerbib\@empty
	\bibitem [{\citenamefont {Xu}\ \emph {et~al.}(2004)\citenamefont {Xu},
		\citenamefont {Lohwongwatana}, \citenamefont {Duan}, \citenamefont
		{Johnson},\ and\ \citenamefont {Garland}}]{Xu_2004}%
	\BibitemOpen
	\bibfield  {author} {\bibinfo {author} {\bibfnamefont {D.}~\bibnamefont
			{Xu}}, \bibinfo {author} {\bibfnamefont {B.}~\bibnamefont {Lohwongwatana}},
		\bibinfo {author} {\bibfnamefont {G.}~\bibnamefont {Duan}}, \bibinfo {author}
		{\bibfnamefont {W.~L.}\ \bibnamefont {Johnson}},\ and\ \bibinfo {author}
		{\bibfnamefont {C.}~\bibnamefont {Garland}},\ }\href
	{https://doi.org/https://doi.org/10.1016/j.actamat.2004.02.009} {\bibfield
		{journal} {\bibinfo  {journal} {Acta Materialia}\ }\textbf {\bibinfo {volume}
			{52}},\ \bibinfo {pages} {2621} (\bibinfo {year} {2004})}\BibitemShut
	{NoStop}%
	\bibitem [{\citenamefont {Wang}\ \emph {et~al.}(2004)\citenamefont {Wang},
		\citenamefont {Li}, \citenamefont {Sun}, \citenamefont {Sui}, \citenamefont
		{Lu},\ and\ \citenamefont {Ma}}]{Wang_2004}%
	\BibitemOpen
	\bibfield  {author} {\bibinfo {author} {\bibfnamefont {D.}~\bibnamefont
			{Wang}}, \bibinfo {author} {\bibfnamefont {Y.}~\bibnamefont {Li}}, \bibinfo
		{author} {\bibfnamefont {B.~B.}\ \bibnamefont {Sun}}, \bibinfo {author}
		{\bibfnamefont {M.~L.}\ \bibnamefont {Sui}}, \bibinfo {author} {\bibfnamefont
			{K.}~\bibnamefont {Lu}},\ and\ \bibinfo {author} {\bibfnamefont
			{E.}~\bibnamefont {Ma}},\ }\href {https://doi.org/10.1063/1.1751219}
	{\bibfield  {journal} {\bibinfo  {journal} {Applied Physics Letters}\
		}\textbf {\bibinfo {volume} {84}},\ \bibinfo {pages} {4029} (\bibinfo {year}
		{2004})},\ \Eprint
	{https://arxiv.org/abs/https://pubs.aip.org/aip/apl/article-pdf/84/20/4029/18588923/4029\_1\_online.pdf}
	{https://pubs.aip.org/aip/apl/article-pdf/84/20/4029/18588923/4029\_1\_online.pdf}
	\BibitemShut {NoStop}%
	\bibitem [{\citenamefont {Xia}\ \emph {et~al.}(2006{\natexlab{a}})\citenamefont
		{Xia}, \citenamefont {Ding}, \citenamefont {Shan},\ and\ \citenamefont
		{Dong}}]{Xia_2006a}%
	\BibitemOpen
	\bibfield  {author} {\bibinfo {author} {\bibfnamefont {L.}~\bibnamefont
			{Xia}}, \bibinfo {author} {\bibfnamefont {D.}~\bibnamefont {Ding}}, \bibinfo
		{author} {\bibfnamefont {S.~T.}\ \bibnamefont {Shan}},\ and\ \bibinfo
		{author} {\bibfnamefont {Y.~D.}\ \bibnamefont {Dong}},\ }\href
	{https://doi.org/10.1088/0953-8984/18/15/002} {\bibfield  {journal} {\bibinfo
			{journal} {Journal of Physics: Condensed Matter}\ }\textbf {\bibinfo
			{volume} {18}},\ \bibinfo {pages} {3543} (\bibinfo {year}
		{2006}{\natexlab{a}})}\BibitemShut {NoStop}%
	\bibitem [{\citenamefont {Xia}\ \emph {et~al.}(2006{\natexlab{b}})\citenamefont
		{Xia}, \citenamefont {Li}, \citenamefont {Fang}, \citenamefont {Wei},\ and\
		\citenamefont {Dong}}]{Xia_2006b}%
	\BibitemOpen
	\bibfield  {author} {\bibinfo {author} {\bibfnamefont {L.}~\bibnamefont
			{Xia}}, \bibinfo {author} {\bibfnamefont {W.~H.}\ \bibnamefont {Li}},
		\bibinfo {author} {\bibfnamefont {S.~S.}\ \bibnamefont {Fang}}, \bibinfo
		{author} {\bibfnamefont {B.~C.}\ \bibnamefont {Wei}},\ and\ \bibinfo {author}
		{\bibfnamefont {Y.~D.}\ \bibnamefont {Dong}},\ }\href
	{https://doi.org/10.1063/1.2158130} {\bibfield  {journal} {\bibinfo
			{journal} {Journal of Applied Physics}\ }\textbf {\bibinfo {volume} {99}},\
		\bibinfo {pages} {026103} (\bibinfo {year} {2006}{\natexlab{b}})},\ \Eprint
	{https://arxiv.org/abs/https://pubs.aip.org/aip/jap/article-pdf/doi/10.1063/1.2158130/14965763/026103\_1\_online.pdf}
	{https://pubs.aip.org/aip/jap/article-pdf/doi/10.1063/1.2158130/14965763/026103\_1\_online.pdf}
	\BibitemShut {NoStop}%
	\bibitem [{\citenamefont {Miracle}\ \emph {et~al.}(2010)\citenamefont
		{Miracle}, \citenamefont {Louzguine-Luzgin}, \citenamefont
		{Louzguina-Luzgina},\ and\ \citenamefont {Inoue}}]{Miracle_2010}%
	\BibitemOpen
	\bibfield  {author} {\bibinfo {author} {\bibfnamefont {D.~B.}\ \bibnamefont
			{Miracle}}, \bibinfo {author} {\bibfnamefont {D.~V.}\ \bibnamefont
			{Louzguine-Luzgin}}, \bibinfo {author} {\bibfnamefont {L.~V.}\ \bibnamefont
			{Louzguina-Luzgina}},\ and\ \bibinfo {author} {\bibfnamefont
			{A.}~\bibnamefont {Inoue}},\ }\href
	{https://doi.org/10.1179/095066010X12646898728200} {\bibfield  {journal}
		{\bibinfo  {journal} {International Materials Reviews}\ }\textbf {\bibinfo
			{volume} {55}},\ \bibinfo {pages} {218} (\bibinfo {year} {2010})},\ \Eprint
	{https://arxiv.org/abs/https://doi.org/10.1179/095066010X12646898728200}
	{https://doi.org/10.1179/095066010X12646898728200} \BibitemShut {NoStop}%
	\bibitem [{\citenamefont {Zhang}\ \emph {et~al.}(2001)\citenamefont {Zhang},
		\citenamefont {Kurosaka},\ and\ \citenamefont {Inoue}}]{Zhang2001}%
	\BibitemOpen
	\bibfield  {author} {\bibinfo {author} {\bibfnamefont {T.}~\bibnamefont
			{Zhang}}, \bibinfo {author} {\bibfnamefont {K.}~\bibnamefont {Kurosaka}},\
		and\ \bibinfo {author} {\bibfnamefont {A.}~\bibnamefont {Inoue}},\ }\href
	{https://doi.org/10.2320/matertrans.42.2042} {\bibfield  {journal} {\bibinfo
			{journal} {Mater. Trans.}\ }\textbf {\bibinfo {volume} {42}},\ \bibinfo
		{pages} {2042} (\bibinfo {year} {2001})}\BibitemShut {NoStop}%
	\bibitem [{\citenamefont {Lee}\ \emph {et~al.}(2008)\citenamefont {Lee},
		\citenamefont {Sun}, \citenamefont {Shin}, \citenamefont {Bae},\ and\
		\citenamefont {Lee}}]{Lee2008}%
	\BibitemOpen
	\bibfield  {author} {\bibinfo {author} {\bibfnamefont {D.-M.}\ \bibnamefont
			{Lee}}, \bibinfo {author} {\bibfnamefont {J.-H.}\ \bibnamefont {Sun}},
		\bibinfo {author} {\bibfnamefont {S.-Y.}\ \bibnamefont {Shin}}, \bibinfo
		{author} {\bibfnamefont {J.-C.}\ \bibnamefont {Bae}},\ and\ \bibinfo {author}
		{\bibfnamefont {C.-H.}\ \bibnamefont {Lee}},\ }\href
	{https://doi.org/10.2320/matertrans.MRP2008066} {\bibfield  {journal}
		{\bibinfo  {journal} {Mater. Trans.}\ }\textbf {\bibinfo {volume} {49}},\
		\bibinfo {pages} {1486} (\bibinfo {year} {2008})}\BibitemShut {NoStop}%
	\bibitem [{\citenamefont {Wu}\ \emph {et~al.}(2008)\citenamefont {Wu},
		\citenamefont {Suo}, \citenamefont {Si}, \citenamefont {Meng},\ and\
		\citenamefont {Qiu}}]{Wu2008}%
	\BibitemOpen
	\bibfield  {author} {\bibinfo {author} {\bibfnamefont {X.}~\bibnamefont
			{Wu}}, \bibinfo {author} {\bibfnamefont {Z.}~\bibnamefont {Suo}}, \bibinfo
		{author} {\bibfnamefont {Y.}~\bibnamefont {Si}}, \bibinfo {author}
		{\bibfnamefont {L.}~\bibnamefont {Meng}},\ and\ \bibinfo {author}
		{\bibfnamefont {K.}~\bibnamefont {Qiu}},\ }\href
	{https://doi.org/https://doi.org/10.1016/j.jallcom.2006.11.010} {\bibfield
		{journal} {\bibinfo  {journal} {J. Alloys Compd.}\ }\textbf {\bibinfo
			{volume} {452}},\ \bibinfo {pages} {268} (\bibinfo {year}
		{2008})}\BibitemShut {NoStop}%
	\bibitem [{\citenamefont {Amore}\ \emph
		{et~al.}(2011{\natexlab{a}})\citenamefont {Amore}, \citenamefont {Brillo},
		\citenamefont {Egry},\ and\ \citenamefont {Novakovic}}]{Amore2011a}%
	\BibitemOpen
	\bibfield  {author} {\bibinfo {author} {\bibfnamefont {S.}~\bibnamefont
			{Amore}}, \bibinfo {author} {\bibfnamefont {J.}~\bibnamefont {Brillo}},
		\bibinfo {author} {\bibfnamefont {I.}~\bibnamefont {Egry}},\ and\ \bibinfo
		{author} {\bibfnamefont {R.}~\bibnamefont {Novakovic}},\ }\href
	{https://doi.org/https://doi.org/10.1016/j.apsusc.2011.04.019} {\bibfield
		{journal} {\bibinfo  {journal} {Appl. Surf. Sci.}\ }\textbf {\bibinfo
			{volume} {257}},\ \bibinfo {pages} {7739} (\bibinfo {year}
		{2011}{\natexlab{a}})}\BibitemShut {NoStop}%
	\bibitem [{\citenamefont {Gargarella}\ \emph {et~al.}(2015)\citenamefont
		{Gargarella}, \citenamefont {Pauly}, \citenamefont {{de Oliveira}},
		\citenamefont {Kühn},\ and\ \citenamefont {Eckert}}]{Gargarella2015}%
	\BibitemOpen
	\bibfield  {author} {\bibinfo {author} {\bibfnamefont {P.}~\bibnamefont
			{Gargarella}}, \bibinfo {author} {\bibfnamefont {S.}~\bibnamefont {Pauly}},
		\bibinfo {author} {\bibfnamefont {M.}~\bibnamefont {{de Oliveira}}}, \bibinfo
		{author} {\bibfnamefont {U.}~\bibnamefont {Kühn}},\ and\ \bibinfo {author}
		{\bibfnamefont {J.}~\bibnamefont {Eckert}},\ }\href
	{https://doi.org/https://doi.org/10.1016/j.jallcom.2014.08.197} {\bibfield
		{journal} {\bibinfo  {journal} {J. Alloys Compd.}\ }\textbf {\bibinfo
			{volume} {618}},\ \bibinfo {pages} {413} (\bibinfo {year}
		{2015})}\BibitemShut {NoStop}%
	\bibitem [{\citenamefont {Shang}\ \emph {et~al.}(2023)\citenamefont {Shang},
		\citenamefont {Jakse}, \citenamefont {Guan}, \citenamefont {Wang},\ and\
		\citenamefont {louis Barrat}}]{Shang2023}%
	\BibitemOpen
	\bibfield  {author} {\bibinfo {author} {\bibfnamefont {B.}~\bibnamefont
			{Shang}}, \bibinfo {author} {\bibfnamefont {N.}~\bibnamefont {Jakse}},
		\bibinfo {author} {\bibfnamefont {P.}~\bibnamefont {Guan}}, \bibinfo {author}
		{\bibfnamefont {W.}~\bibnamefont {Wang}},\ and\ \bibinfo {author}
		{\bibfnamefont {J.}~\bibnamefont {louis Barrat}},\ }\href
	{https://doi.org/https://doi.org/10.1016/j.actamat.2022.118668} {\bibfield
		{journal} {\bibinfo  {journal} {Acta Materialia}\ }\textbf {\bibinfo {volume}
			{246}},\ \bibinfo {pages} {118668} (\bibinfo {year} {2023})}\BibitemShut
	{NoStop}%
	\bibitem [{\citenamefont {Cao}\ \emph {et~al.}(2024)\citenamefont {Cao},
		\citenamefont {Xia}, \citenamefont {Jakse}, \citenamefont {Zeng},
		\citenamefont {Yu}, \citenamefont {Zhao}, \citenamefont {Lu},\ and\
		\citenamefont {Li}}]{Cao2024}%
	\BibitemOpen
	\bibfield  {author} {\bibinfo {author} {\bibfnamefont {S.}~\bibnamefont
			{Cao}}, \bibinfo {author} {\bibfnamefont {M.}~\bibnamefont {Xia}}, \bibinfo
		{author} {\bibfnamefont {N.}~\bibnamefont {Jakse}}, \bibinfo {author}
		{\bibfnamefont {L.}~\bibnamefont {Zeng}}, \bibinfo {author} {\bibfnamefont
			{P.}~\bibnamefont {Yu}}, \bibinfo {author} {\bibfnamefont {Y.}~\bibnamefont
			{Zhao}}, \bibinfo {author} {\bibfnamefont {W.}~\bibnamefont {Lu}},\ and\
		\bibinfo {author} {\bibfnamefont {J.}~\bibnamefont {Li}},\ }\href
	{https://doi.org/https://doi.org/10.1016/j.scriptamat.2024.116143} {\bibfield
		{journal} {\bibinfo  {journal} {Scripta Materialia}\ }\textbf {\bibinfo
			{volume} {248}},\ \bibinfo {pages} {116143} (\bibinfo {year}
		{2024})}\BibitemShut {NoStop}%
	\bibitem [{\citenamefont {Mei-Bo}\ \emph {et~al.}(2004)\citenamefont {Mei-Bo},
		\citenamefont {De-Qian}, \citenamefont {Ming-Xiang},\ and\ \citenamefont
		{Wei-Hua}}]{Mei-Bo2004}%
	\BibitemOpen
	\bibfield  {author} {\bibinfo {author} {\bibfnamefont {T.}~\bibnamefont
			{Mei-Bo}}, \bibinfo {author} {\bibfnamefont {Z.}~\bibnamefont {De-Qian}},
		\bibinfo {author} {\bibfnamefont {P.}~\bibnamefont {Ming-Xiang}},\ and\
		\bibinfo {author} {\bibfnamefont {W.}~\bibnamefont {Wei-Hua}},\ }\href@noop
	{} {\bibfield  {journal} {\bibinfo  {journal} {Chinese Physics Letters}\
		}\textbf {\bibinfo {volume} {21}},\ \bibinfo {pages} {901} (\bibinfo {year}
		{2004})}\BibitemShut {NoStop}%
	\bibitem [{\citenamefont {Jakse}\ and\ \citenamefont
		{Pasturel}(2008{\natexlab{a}})}]{Jakse_2008a}%
	\BibitemOpen
	\bibfield  {author} {\bibinfo {author} {\bibfnamefont {N.}~\bibnamefont
			{Jakse}}\ and\ \bibinfo {author} {\bibfnamefont {A.}~\bibnamefont
			{Pasturel}},\ }\href {https://doi.org/10.1103/PhysRevB.78.214204} {\bibfield
		{journal} {\bibinfo  {journal} {Phys. Rev. B}\ }\textbf {\bibinfo {volume}
			{78}},\ \bibinfo {pages} {214204} (\bibinfo {year}
		{2008}{\natexlab{a}})}\BibitemShut {NoStop}%
	\bibitem [{\citenamefont {{Yang, F.}}\ \emph {et~al.}(2014)\citenamefont
		{{Yang, F.}}, \citenamefont {{Holland-Moritz, D.}}, \citenamefont {{Gegner,
				J.}}, \citenamefont {{Heintzmann, P.}}, \citenamefont {{Kargl, F.}},
		\citenamefont {{Yuan, C. C.}}, \citenamefont {{Simeoni, G. G.}},\ and\
		\citenamefont {{Meyer, A.}}}]{Yang2014}%
	\BibitemOpen
	\bibfield  {author} {\bibinfo {author} {\bibnamefont {{Yang, F.}}}, \bibinfo
		{author} {\bibnamefont {{Holland-Moritz, D.}}}, \bibinfo {author}
		{\bibnamefont {{Gegner, J.}}}, \bibinfo {author} {\bibnamefont {{Heintzmann,
					P.}}}, \bibinfo {author} {\bibnamefont {{Kargl, F.}}}, \bibinfo {author}
		{\bibnamefont {{Yuan, C. C.}}}, \bibinfo {author} {\bibnamefont {{Simeoni, G.
					G.}}},\ and\ \bibinfo {author} {\bibnamefont {{Meyer, A.}}},\ }\href
	{https://doi.org/10.1209/0295-5075/107/46001} {\bibfield  {journal} {\bibinfo
			{journal} {Europhys. Lett.}\ }\textbf {\bibinfo {volume} {107}},\ \bibinfo
		{pages} {46001} (\bibinfo {year} {2014})}\BibitemShut {NoStop}%
	\bibitem [{\citenamefont {Jakse}\ \emph {et~al.}(2013)\citenamefont {Jakse},
		\citenamefont {Nguyen},\ and\ \citenamefont {Pasturel}}]{Jakse2013b}%
	\BibitemOpen
	\bibfield  {author} {\bibinfo {author} {\bibfnamefont {N.}~\bibnamefont
			{Jakse}}, \bibinfo {author} {\bibfnamefont {T.~L.~T.}\ \bibnamefont
			{Nguyen}},\ and\ \bibinfo {author} {\bibfnamefont {A.}~\bibnamefont
			{Pasturel}},\ }\href {https://doi.org/10.1063/1.4817426} {\bibfield
		{journal} {\bibinfo  {journal} {Journal of Applied Physics}\ }\textbf
		{\bibinfo {volume} {114}},\ \bibinfo {pages} {063514} (\bibinfo {year}
		{2013})},\ \Eprint
	{https://arxiv.org/abs/https://pubs.aip.org/aip/jap/article-pdf/doi/10.1063/1.4817426/15116540/063514\_1\_online.pdf}
	{https://pubs.aip.org/aip/jap/article-pdf/doi/10.1063/1.4817426/15116540/063514\_1\_online.pdf}
	\BibitemShut {NoStop}%
	\bibitem [{\citenamefont {Holland-Moritz}\ \emph {et~al.}(2009)\citenamefont
		{Holland-Moritz}, \citenamefont {St\"uber}, \citenamefont {Hartmann},
		\citenamefont {Unruh}, \citenamefont {Hansen},\ and\ \citenamefont
		{Meyer}}]{Holland-Moritz2009}%
	\BibitemOpen
	\bibfield  {author} {\bibinfo {author} {\bibfnamefont {D.}~\bibnamefont
			{Holland-Moritz}}, \bibinfo {author} {\bibfnamefont {S.}~\bibnamefont
			{St\"uber}}, \bibinfo {author} {\bibfnamefont {H.}~\bibnamefont {Hartmann}},
		\bibinfo {author} {\bibfnamefont {T.}~\bibnamefont {Unruh}}, \bibinfo
		{author} {\bibfnamefont {T.}~\bibnamefont {Hansen}},\ and\ \bibinfo {author}
		{\bibfnamefont {A.}~\bibnamefont {Meyer}},\ }\href
	{https://doi.org/10.1103/PhysRevB.79.064204} {\bibfield  {journal} {\bibinfo
			{journal} {Phys. Rev. B}\ }\textbf {\bibinfo {volume} {79}},\ \bibinfo
		{pages} {064204} (\bibinfo {year} {2009})}\BibitemShut {NoStop}%
	\bibitem [{\citenamefont {Holland-Moritz}\ \emph {et~al.}(2012)\citenamefont
		{Holland-Moritz}, \citenamefont {Yang}, \citenamefont {Kordel}, \citenamefont
		{Klein}, \citenamefont {Kargl}, \citenamefont {Gegner}, \citenamefont
		{Hansen}, \citenamefont {Bednarcik}, \citenamefont {Kaban}, \citenamefont
		{Shuleshova}, \citenamefont {Mattern},\ and\ \citenamefont
		{Meyer}}]{Holland-Moritz_2012}%
	\BibitemOpen
	\bibfield  {author} {\bibinfo {author} {\bibfnamefont {D.}~\bibnamefont
			{Holland-Moritz}}, \bibinfo {author} {\bibfnamefont {F.}~\bibnamefont
			{Yang}}, \bibinfo {author} {\bibfnamefont {T.}~\bibnamefont {Kordel}},
		\bibinfo {author} {\bibfnamefont {S.}~\bibnamefont {Klein}}, \bibinfo
		{author} {\bibfnamefont {F.}~\bibnamefont {Kargl}}, \bibinfo {author}
		{\bibfnamefont {J.}~\bibnamefont {Gegner}}, \bibinfo {author} {\bibfnamefont
			{T.}~\bibnamefont {Hansen}}, \bibinfo {author} {\bibfnamefont
			{J.}~\bibnamefont {Bednarcik}}, \bibinfo {author} {\bibfnamefont
			{I.}~\bibnamefont {Kaban}}, \bibinfo {author} {\bibfnamefont
			{O.}~\bibnamefont {Shuleshova}}, \bibinfo {author} {\bibfnamefont
			{N.}~\bibnamefont {Mattern}},\ and\ \bibinfo {author} {\bibfnamefont
			{A.}~\bibnamefont {Meyer}},\ }\href
	{https://doi.org/10.1209/0295-5075/100/56002} {\bibfield  {journal} {\bibinfo
			{journal} {Europhysics Letters}\ }\textbf {\bibinfo {volume} {100}},\
		\bibinfo {pages} {56002} (\bibinfo {year} {2012})}\BibitemShut {NoStop}%
	\bibitem [{\citenamefont {Herlach}\ \emph {et~al.}(2007)\citenamefont
		{Herlach}, \citenamefont {Galenko},\ and\ \citenamefont
		{Holland-Moritz}}]{Herlach_2007}%
	\BibitemOpen
	\bibfield  {author} {\bibinfo {author} {\bibfnamefont {D.}~\bibnamefont
			{Herlach}}, \bibinfo {author} {\bibfnamefont {P.}~\bibnamefont {Galenko}},\
		and\ \bibinfo {author} {\bibfnamefont {D.}~\bibnamefont {Holland-Moritz}},\
	}\href {https://cds.cern.ch/record/1034906} {\emph {\bibinfo {title}
			{{Metastable Solids from Undercooled Melts}}}}\ (\bibinfo  {publisher}
	{Elsevier},\ \bibinfo {address} {Burlington, MA},\ \bibinfo {year}
	{2007})\BibitemShut {NoStop}%
	\bibitem [{\citenamefont {Nowak}\ \emph {et~al.}(2017)\citenamefont {Nowak},
		\citenamefont {Holland-Moritz}, \citenamefont {Yang}, \citenamefont
		{Voigtmann}, \citenamefont {Evenson}, \citenamefont {Hansen},\ and\
		\citenamefont {Meyer}}]{Nowak2017a}%
	\BibitemOpen
	\bibfield  {author} {\bibinfo {author} {\bibfnamefont {B.}~\bibnamefont
			{Nowak}}, \bibinfo {author} {\bibfnamefont {D.}~\bibnamefont
			{Holland-Moritz}}, \bibinfo {author} {\bibfnamefont {F.}~\bibnamefont
			{Yang}}, \bibinfo {author} {\bibfnamefont {T.}~\bibnamefont {Voigtmann}},
		\bibinfo {author} {\bibfnamefont {Z.}~\bibnamefont {Evenson}}, \bibinfo
		{author} {\bibfnamefont {T.~C.}\ \bibnamefont {Hansen}},\ and\ \bibinfo
		{author} {\bibfnamefont {A.}~\bibnamefont {Meyer}},\ }\href
	{https://doi.org/10.1103/PhysRevB.96.054201} {\bibfield  {journal} {\bibinfo
			{journal} {Phys. Rev. B}\ }\textbf {\bibinfo {volume} {96}},\ \bibinfo
		{pages} {054201} (\bibinfo {year} {2017})}\BibitemShut {NoStop}%
	\bibitem [{\citenamefont {Kreuzer}\ \emph {et~al.}(2024)\citenamefont
		{Kreuzer}, \citenamefont {Yang}, \citenamefont {Evenson}, \citenamefont
		{Holland-Moritz}, \citenamefont {Bernasconi}, \citenamefont {Hansen},
		\citenamefont {Blankenburg}, \citenamefont {Meyer},\ and\ \citenamefont
		{Petry}}]{Kreuzer_2024}%
	\BibitemOpen
	\bibfield  {author} {\bibinfo {author} {\bibfnamefont {L.~P.}\ \bibnamefont
			{Kreuzer}}, \bibinfo {author} {\bibfnamefont {F.}~\bibnamefont {Yang}},
		\bibinfo {author} {\bibfnamefont {Z.}~\bibnamefont {Evenson}}, \bibinfo
		{author} {\bibfnamefont {D.}~\bibnamefont {Holland-Moritz}}, \bibinfo
		{author} {\bibfnamefont {A.}~\bibnamefont {Bernasconi}}, \bibinfo {author}
		{\bibfnamefont {T.~C.}\ \bibnamefont {Hansen}}, \bibinfo {author}
		{\bibfnamefont {M.}~\bibnamefont {Blankenburg}}, \bibinfo {author}
		{\bibfnamefont {A.}~\bibnamefont {Meyer}},\ and\ \bibinfo {author}
		{\bibfnamefont {W.}~\bibnamefont {Petry}},\ }\href
	{https://doi.org/10.1103/PhysRevB.109.174108} {\bibfield  {journal} {\bibinfo
			{journal} {Phys. Rev. B}\ }\textbf {\bibinfo {volume} {109}},\ \bibinfo
		{pages} {174108} (\bibinfo {year} {2024})}\BibitemShut {NoStop}%
	\bibitem [{\citenamefont {Jakse}\ and\ \citenamefont
		{Pasturel}(2008{\natexlab{b}})}]{Jakse_2008b}%
	\BibitemOpen
	\bibfield  {author} {\bibinfo {author} {\bibfnamefont {N.}~\bibnamefont
			{Jakse}}\ and\ \bibinfo {author} {\bibfnamefont {A.}~\bibnamefont
			{Pasturel}},\ }\href {https://doi.org/10.1063/1.2976428} {\bibfield
		{journal} {\bibinfo  {journal} {Applied Physics Letters}\ }\textbf {\bibinfo
			{volume} {93}},\ \bibinfo {pages} {113104} (\bibinfo {year}
		{2008}{\natexlab{b}})},\ \Eprint
	{https://arxiv.org/abs/https://pubs.aip.org/aip/apl/article-pdf/doi/10.1063/1.2976428/14402575/113104\_1\_online.pdf}
	{https://pubs.aip.org/aip/apl/article-pdf/doi/10.1063/1.2976428/14402575/113104\_1\_online.pdf}
	\BibitemShut {NoStop}%
	\bibitem [{\citenamefont {Woodward}\ \emph {et~al.}(2010)\citenamefont
		{Woodward}, \citenamefont {Asta}, \citenamefont {Trinkle}, \citenamefont
		{Lill},\ and\ \citenamefont {Angioletti-Uberti}}]{Woodward_2010}%
	\BibitemOpen
	\bibfield  {author} {\bibinfo {author} {\bibfnamefont {C.}~\bibnamefont
			{Woodward}}, \bibinfo {author} {\bibfnamefont {M.}~\bibnamefont {Asta}},
		\bibinfo {author} {\bibfnamefont {D.~R.}\ \bibnamefont {Trinkle}}, \bibinfo
		{author} {\bibfnamefont {J.}~\bibnamefont {Lill}},\ and\ \bibinfo {author}
		{\bibfnamefont {S.}~\bibnamefont {Angioletti-Uberti}},\ }\href
	{https://doi.org/10.1063/1.3437644} {\bibfield  {journal} {\bibinfo
			{journal} {Journal of Applied Physics}\ }\textbf {\bibinfo {volume} {107}},\
		\bibinfo {pages} {113522} (\bibinfo {year} {2010})},\ \Eprint
	{https://arxiv.org/abs/https://pubs.aip.org/aip/jap/article-pdf/doi/10.1063/1.3437644/13197405/113522\_1\_online.pdf}
	{https://pubs.aip.org/aip/jap/article-pdf/doi/10.1063/1.3437644/13197405/113522\_1\_online.pdf}
	\BibitemShut {NoStop}%
	\bibitem [{\citenamefont {Jakse}\ and\ \citenamefont
		{Pasturel}(2016)}]{Jakse_2016}%
	\BibitemOpen
	\bibfield  {author} {\bibinfo {author} {\bibfnamefont {N.}~\bibnamefont
			{Jakse}}\ and\ \bibinfo {author} {\bibfnamefont {A.}~\bibnamefont
			{Pasturel}},\ }\href {https://doi.org/10.1103/PhysRevB.94.224201} {\bibfield
		{journal} {\bibinfo  {journal} {Phys. Rev. B}\ }\textbf {\bibinfo {volume}
			{94}},\ \bibinfo {pages} {224201} (\bibinfo {year} {2016})}\BibitemShut
	{NoStop}%
	\bibitem [{\citenamefont {Pasturel}\ and\ \citenamefont
		{Jakse}(2016)}]{Pasturel_2016}%
	\BibitemOpen
	\bibfield  {author} {\bibinfo {author} {\bibfnamefont {A.}~\bibnamefont
			{Pasturel}}\ and\ \bibinfo {author} {\bibfnamefont {N.}~\bibnamefont
			{Jakse}},\ }\href {https://doi.org/10.1063/1.4960015} {\bibfield  {journal}
		{\bibinfo  {journal} {Applied Physics Letters}\ }\textbf {\bibinfo {volume}
			{109}},\ \bibinfo {pages} {041904} (\bibinfo {year} {2016})},\ \Eprint
	{https://arxiv.org/abs/https://pubs.aip.org/aip/apl/article-pdf/doi/10.1063/1.4960015/14482670/041904\_1\_online.pdf}
	{https://pubs.aip.org/aip/apl/article-pdf/doi/10.1063/1.4960015/14482670/041904\_1\_online.pdf}
	\BibitemShut {NoStop}%
	\bibitem [{\citenamefont {Li}\ \emph {et~al.}(2023)\citenamefont {Li},
		\citenamefont {Xiao}, \citenamefont {Qin}, \citenamefont {Ruan},\ and\
		\citenamefont {Li}}]{Li_2023}%
	\BibitemOpen
	\bibfield  {author} {\bibinfo {author} {\bibfnamefont {J.}~\bibnamefont
			{Li}}, \bibinfo {author} {\bibfnamefont {R.}~\bibnamefont {Xiao}}, \bibinfo
		{author} {\bibfnamefont {J.}~\bibnamefont {Qin}}, \bibinfo {author}
		{\bibfnamefont {Y.}~\bibnamefont {Ruan}},\ and\ \bibinfo {author}
		{\bibfnamefont {H.}~\bibnamefont {Li}},\ }\href
	{https://doi.org/https://doi.org/10.1016/j.commatsci.2023.112499} {\bibfield
		{journal} {\bibinfo  {journal} {Computational Materials Science}\ }\textbf
		{\bibinfo {volume} {230}},\ \bibinfo {pages} {112499} (\bibinfo {year}
		{2023})}\BibitemShut {NoStop}%
	\bibitem [{\citenamefont {Kresse}\ and\ \citenamefont
		{Hafner}(1993)}]{Kresse_1993}%
	\BibitemOpen
	\bibfield  {author} {\bibinfo {author} {\bibfnamefont {G.}~\bibnamefont
			{Kresse}}\ and\ \bibinfo {author} {\bibfnamefont {J.}~\bibnamefont
			{Hafner}},\ }\href {https://doi.org/10.1103/PhysRevB.47.558} {\bibfield
		{journal} {\bibinfo  {journal} {Phys. Rev. B}\ }\textbf {\bibinfo {volume}
			{47}},\ \bibinfo {pages} {558} (\bibinfo {year} {1993})}\BibitemShut
	{NoStop}%
	\bibitem [{\citenamefont {Bl\"ochl}(1994)}]{Blochl1994}%
	\BibitemOpen
	\bibfield  {author} {\bibinfo {author} {\bibfnamefont {P.~E.}\ \bibnamefont
			{Bl\"ochl}},\ }\href {https://doi.org/10.1103/PhysRevB.50.17953} {\bibfield
		{journal} {\bibinfo  {journal} {Phys. Rev. B}\ }\textbf {\bibinfo {volume}
			{50}},\ \bibinfo {pages} {17953} (\bibinfo {year} {1994})}\BibitemShut
	{NoStop}%
	\bibitem [{\citenamefont {Kresse}\ and\ \citenamefont
		{Joubert}(1999)}]{Kresse1999}%
	\BibitemOpen
	\bibfield  {author} {\bibinfo {author} {\bibfnamefont {G.}~\bibnamefont
			{Kresse}}\ and\ \bibinfo {author} {\bibfnamefont {D.}~\bibnamefont
			{Joubert}},\ }\href {https://doi.org/10.1103/PhysRevB.59.1758} {\bibfield
		{journal} {\bibinfo  {journal} {Phys. Rev. B}\ }\textbf {\bibinfo {volume}
			{59}},\ \bibinfo {pages} {1758} (\bibinfo {year} {1999})}\BibitemShut
	{NoStop}%
	\bibitem [{\citenamefont {Perdew}\ and\ \citenamefont
		{Zunger}(1981)}]{Perdew1981}%
	\BibitemOpen
	\bibfield  {author} {\bibinfo {author} {\bibfnamefont {J.~P.}\ \bibnamefont
			{Perdew}}\ and\ \bibinfo {author} {\bibfnamefont {A.}~\bibnamefont
			{Zunger}},\ }\href {https://doi.org/10.1103/PhysRevB.23.5048} {\bibfield
		{journal} {\bibinfo  {journal} {Phys. Rev. B}\ }\textbf {\bibinfo {volume}
			{23}},\ \bibinfo {pages} {5048} (\bibinfo {year} {1981})}\BibitemShut
	{NoStop}%
	\bibitem [{\citenamefont {Nos{\'e}}(1984)}]{Nose_1984}%
	\BibitemOpen
	\bibfield  {author} {\bibinfo {author} {\bibfnamefont {S.}~\bibnamefont
			{Nos{\'e}}},\ }\href@noop {} {\bibfield  {journal} {\bibinfo  {journal} {The
				Journal of chemical physics}\ }\textbf {\bibinfo {volume} {81}},\ \bibinfo
		{pages} {511} (\bibinfo {year} {1984})}\BibitemShut {NoStop}%
	\bibitem [{\citenamefont {Binder}\ and\ \citenamefont
		{Kob}(2011)}]{Binder_2011}%
	\BibitemOpen
	\bibfield  {author} {\bibinfo {author} {\bibfnamefont {K.}~\bibnamefont
			{Binder}}\ and\ \bibinfo {author} {\bibfnamefont {W.}~\bibnamefont {Kob}},\
	}\href {https://doi.org/10.1142/7300} {\emph {\bibinfo {title} {Glassy
				Materials and Disordered Solids}}},\ \bibinfo {edition} {revised}\ ed.\
	(\bibinfo  {publisher} {WORLD SCIENTIFIC},\ \bibinfo {year} {2011})\ \Eprint
	{https://arxiv.org/abs/https://www.worldscientific.com/doi/pdf/10.1142/7300}
	{https://www.worldscientific.com/doi/pdf/10.1142/7300} \BibitemShut {NoStop}%
	\bibitem [{\citenamefont {Stillinger}\ and\ \citenamefont
		{Weber}(1982)}]{Stillinger_1982}%
	\BibitemOpen
	\bibfield  {author} {\bibinfo {author} {\bibfnamefont {F.~H.}\ \bibnamefont
			{Stillinger}}\ and\ \bibinfo {author} {\bibfnamefont {T.~A.}\ \bibnamefont
			{Weber}},\ }\href@noop {} {\bibfield  {journal} {\bibinfo  {journal}
			{Physical Review A}\ }\textbf {\bibinfo {volume} {25}},\ \bibinfo {pages}
		{978} (\bibinfo {year} {1982})}\BibitemShut {NoStop}%
	\bibitem [{\citenamefont {Faken}\ and\ \citenamefont
		{Jónsson}(1994)}]{Faken_1994}%
	\BibitemOpen
	\bibfield  {author} {\bibinfo {author} {\bibfnamefont {D.}~\bibnamefont
			{Faken}}\ and\ \bibinfo {author} {\bibfnamefont {H.}~\bibnamefont
			{Jónsson}},\ }\href
	{https://doi.org/https://doi.org/10.1016/0927-0256(94)90109-0} {\bibfield
		{journal} {\bibinfo  {journal} {Computational Materials Science}\ }\textbf
		{\bibinfo {volume} {2}},\ \bibinfo {pages} {279} (\bibinfo {year}
		{1994})}\BibitemShut {NoStop}%
	\bibitem [{\citenamefont {Stukowski}(2009)}]{Stukowski_2010}%
	\BibitemOpen
	\bibfield  {author} {\bibinfo {author} {\bibfnamefont {A.}~\bibnamefont
			{Stukowski}},\ }\href {https://doi.org/10.1088/0965-0393/18/1/015012}
	{\bibfield  {journal} {\bibinfo  {journal} {Modelling and Simulation in
				Materials Science and Engineering}\ }\textbf {\bibinfo {volume} {18}},\
		\bibinfo {pages} {015012} (\bibinfo {year} {2009})}\BibitemShut {NoStop}%
	\bibitem [{\citenamefont {Holland-Moritz}\ \emph {et~al.}(2023)\citenamefont
		{Holland-Moritz}, \citenamefont {Yang}, \citenamefont {Hansen},\ and\
		\citenamefont {Kargl}}]{Holland-Moritz_2023}%
	\BibitemOpen
	\bibfield  {author} {\bibinfo {author} {\bibfnamefont {D.}~\bibnamefont
			{Holland-Moritz}}, \bibinfo {author} {\bibfnamefont {F.}~\bibnamefont
			{Yang}}, \bibinfo {author} {\bibfnamefont {T.}~\bibnamefont {Hansen}},\ and\
		\bibinfo {author} {\bibfnamefont {F.}~\bibnamefont {Kargl}},\ }\href
	{https://doi.org/10.1088/1361-648X/aceee0} {\bibfield  {journal} {\bibinfo
			{journal} {Journal of Physics: Condensed Matter}\ }\textbf {\bibinfo {volume}
			{35}} (\bibinfo {year} {2023})}\BibitemShut {NoStop}%
	\bibitem [{\citenamefont {Holland-Moritz}\ \emph {et~al.}(2007)\citenamefont
		{Holland-Moritz}, \citenamefont {Heinen}, \citenamefont {Bellissent},\ and\
		\citenamefont {Schenk}}]{Holland-Moritz2007}%
	\BibitemOpen
	\bibfield  {author} {\bibinfo {author} {\bibfnamefont {D.}~\bibnamefont
			{Holland-Moritz}}, \bibinfo {author} {\bibfnamefont {O.}~\bibnamefont
			{Heinen}}, \bibinfo {author} {\bibfnamefont {R.}~\bibnamefont {Bellissent}},\
		and\ \bibinfo {author} {\bibfnamefont {T.}~\bibnamefont {Schenk}},\ }\href
	{https://doi.org/https://doi.org/10.1016/j.msea.2005.12.093} {\bibfield
		{journal} {\bibinfo  {journal} {Mater. Sci. Eng. A}\ }\textbf {\bibinfo
			{volume} {449-451}},\ \bibinfo {pages} {42} (\bibinfo {year}
		{2007})}\BibitemShut {NoStop}%
	\bibitem [{\citenamefont {Sachdev}\ and\ \citenamefont
		{Nelson}(1984)}]{Sachdev1984}%
	\BibitemOpen
	\bibfield  {author} {\bibinfo {author} {\bibfnamefont {S.}~\bibnamefont
			{Sachdev}}\ and\ \bibinfo {author} {\bibfnamefont {D.~R.}\ \bibnamefont
			{Nelson}},\ }\href {https://doi.org/10.1103/PhysRevLett.53.1947} {\bibfield
		{journal} {\bibinfo  {journal} {Phys. Rev. Lett.}\ }\textbf {\bibinfo
			{volume} {53}},\ \bibinfo {pages} {1947} (\bibinfo {year}
		{1984})}\BibitemShut {NoStop}%
	\bibitem [{\citenamefont {Schenk}\ \emph {et~al.}(2002)\citenamefont {Schenk},
		\citenamefont {Holland-Moritz}, \citenamefont {Simonet}, \citenamefont
		{Bellissent},\ and\ \citenamefont {Herlach}}]{Schenk_2002}%
	\BibitemOpen
	\bibfield  {author} {\bibinfo {author} {\bibfnamefont {T.}~\bibnamefont
			{Schenk}}, \bibinfo {author} {\bibfnamefont {D.}~\bibnamefont
			{Holland-Moritz}}, \bibinfo {author} {\bibfnamefont {V.}~\bibnamefont
			{Simonet}}, \bibinfo {author} {\bibfnamefont {R.}~\bibnamefont
			{Bellissent}},\ and\ \bibinfo {author} {\bibfnamefont {D.~M.}\ \bibnamefont
			{Herlach}},\ }\href {https://doi.org/10.1103/PhysRevLett.89.075507}
	{\bibfield  {journal} {\bibinfo  {journal} {Phys. Rev. Lett.}\ }\textbf
		{\bibinfo {volume} {89}},\ \bibinfo {pages} {075507} (\bibinfo {year}
		{2002})}\BibitemShut {NoStop}%
	\bibitem [{\citenamefont {J.-P.}\ and\ \citenamefont
		{I.R.}(2006)}]{Hansen2006}%
	\BibitemOpen
	\bibfield  {author} {\bibinfo {author} {\bibfnamefont {H.}~\bibnamefont
			{J.-P.}}\ and\ \bibinfo {author} {\bibfnamefont {M.}~\bibnamefont {I.R.}},\
	}\href@noop {} {\emph {\bibinfo {title} {The Theory of Simple Liquids}}},\
	3rd edition\ (\bibinfo  {publisher} {Academic Press, Oxford},\ \bibinfo
	{year} {2006})\BibitemShut {NoStop}%
	\bibitem [{\citenamefont {Amore}\ \emph
		{et~al.}(2011{\natexlab{b}})\citenamefont {Amore}, \citenamefont {Horbach},\
		and\ \citenamefont {Egry}}]{Amore2011b}%
	\BibitemOpen
	\bibfield  {author} {\bibinfo {author} {\bibfnamefont {S.}~\bibnamefont
			{Amore}}, \bibinfo {author} {\bibfnamefont {J.}~\bibnamefont {Horbach}},\
		and\ \bibinfo {author} {\bibfnamefont {I.}~\bibnamefont {Egry}},\ }\href
	{https://doi.org/10.1063/1.3528217} {\bibfield  {journal} {\bibinfo
			{journal} {J. Chem. Phys.}\ }\textbf {\bibinfo {volume} {134}},\ \bibinfo
		{pages} {044515} (\bibinfo {year} {2011}{\natexlab{b}})}\BibitemShut
	{NoStop}%
	\bibitem [{\citenamefont {Amore}\ \emph {et~al.}(2013)\citenamefont {Amore},
		\citenamefont {Delsante}, \citenamefont {Kobatake},\ and\ \citenamefont
		{Brillo}}]{Amore2013}%
	\BibitemOpen
	\bibfield  {author} {\bibinfo {author} {\bibfnamefont {S.}~\bibnamefont
			{Amore}}, \bibinfo {author} {\bibfnamefont {S.}~\bibnamefont {Delsante}},
		\bibinfo {author} {\bibfnamefont {H.}~\bibnamefont {Kobatake}},\ and\
		\bibinfo {author} {\bibfnamefont {J.}~\bibnamefont {Brillo}},\ }\href
	{https://doi.org/10.1063/1.4817679} {\bibfield  {journal} {\bibinfo
			{journal} {J. Chem. Phys.}\ }\textbf {\bibinfo {volume} {139}},\ \bibinfo
		{pages} {064504} (\bibinfo {year} {2013})}\BibitemShut {NoStop}%
	\bibitem [{\citenamefont {Frank}\ and\ \citenamefont
		{Mott}(1952)}]{Frank_1952}%
	\BibitemOpen
	\bibfield  {author} {\bibinfo {author} {\bibfnamefont {F.~C.}\ \bibnamefont
			{Frank}}\ and\ \bibinfo {author} {\bibfnamefont {N.~F.}\ \bibnamefont
			{Mott}},\ }\href {https://doi.org/10.1098/rspa.1952.0194} {\bibfield
		{journal} {\bibinfo  {journal} {Proceedings of the Royal Society of London.
				Series A. Mathematical and Physical Sciences}\ }\textbf {\bibinfo {volume}
			{215}},\ \bibinfo {pages} {43} (\bibinfo {year} {1952})},\ \Eprint
	{https://arxiv.org/abs/https://royalsocietypublishing.org/doi/pdf/10.1098/rspa.1952.0194}
	{https://royalsocietypublishing.org/doi/pdf/10.1098/rspa.1952.0194}
	\BibitemShut {NoStop}%
	\bibitem [{\citenamefont {Jakse}\ and\ \citenamefont
		{Pasturel}(2003)}]{Jakse2003}%
	\BibitemOpen
	\bibfield  {author} {\bibinfo {author} {\bibfnamefont {N.}~\bibnamefont
			{Jakse}}\ and\ \bibinfo {author} {\bibfnamefont {A.}~\bibnamefont
			{Pasturel}},\ }\href {https://doi.org/10.1103/PhysRevLett.91.195501}
	{\bibfield  {journal} {\bibinfo  {journal} {Phys. Rev. Lett.}\ }\textbf
		{\bibinfo {volume} {91}},\ \bibinfo {pages} {195501} (\bibinfo {year}
		{2003})}\BibitemShut {NoStop}%
	\bibitem [{\citenamefont {Basuki}\ \emph {et~al.}(2017)\citenamefont {Basuki},
		\citenamefont {Yang}, \citenamefont {Gill}, \citenamefont {R\"atzke},
		\citenamefont {Meyer},\ and\ \citenamefont {Faupel}}]{Basuki2017}%
	\BibitemOpen
	\bibfield  {author} {\bibinfo {author} {\bibfnamefont {S.~W.}\ \bibnamefont
			{Basuki}}, \bibinfo {author} {\bibfnamefont {F.}~\bibnamefont {Yang}},
		\bibinfo {author} {\bibfnamefont {E.}~\bibnamefont {Gill}}, \bibinfo {author}
		{\bibfnamefont {K.}~\bibnamefont {R\"atzke}}, \bibinfo {author}
		{\bibfnamefont {A.}~\bibnamefont {Meyer}},\ and\ \bibinfo {author}
		{\bibfnamefont {F.}~\bibnamefont {Faupel}},\ }\href
	{https://doi.org/10.1103/PhysRevB.95.024301} {\bibfield  {journal} {\bibinfo
			{journal} {Phys. Rev. B}\ }\textbf {\bibinfo {volume} {95}},\ \bibinfo
		{pages} {024301} (\bibinfo {year} {2017})}\BibitemShut {NoStop}%
	\bibitem [{\citenamefont {Jakse}\ and\ \citenamefont
		{Pasturel}(2013)}]{Jakse_2013}%
	\BibitemOpen
	\bibfield  {author} {\bibinfo {author} {\bibfnamefont {N.}~\bibnamefont
			{Jakse}}\ and\ \bibinfo {author} {\bibfnamefont {A.}~\bibnamefont
			{Pasturel}},\ }\href {https://doi.org/10.1038/srep03135} {\bibfield
		{journal} {\bibinfo  {journal} {Scientific Reports}\ }\textbf {\bibinfo
			{volume} {3}},\ \bibinfo {pages} {3135} (\bibinfo {year} {2013})}\BibitemShut
	{NoStop}%
	\bibitem [{\citenamefont {Assael}\ \emph {et~al.}(2010)\citenamefont {Assael},
		\citenamefont {Kalyva}, \citenamefont {Antoniadis}, \citenamefont
		{Michael~Banish}, \citenamefont {Egry}, \citenamefont {Wu}, \citenamefont
		{Kaschnitz},\ and\ \citenamefont {Wakeham}}]{Assael2010}%
	\BibitemOpen
	\bibfield  {author} {\bibinfo {author} {\bibfnamefont {M.~J.}\ \bibnamefont
			{Assael}}, \bibinfo {author} {\bibfnamefont {A.~E.}\ \bibnamefont {Kalyva}},
		\bibinfo {author} {\bibfnamefont {K.~D.}\ \bibnamefont {Antoniadis}},
		\bibinfo {author} {\bibfnamefont {R.}~\bibnamefont {Michael~Banish}},
		\bibinfo {author} {\bibfnamefont {I.}~\bibnamefont {Egry}}, \bibinfo {author}
		{\bibfnamefont {J.}~\bibnamefont {Wu}}, \bibinfo {author} {\bibfnamefont
			{E.}~\bibnamefont {Kaschnitz}},\ and\ \bibinfo {author} {\bibfnamefont
			{W.~A.}\ \bibnamefont {Wakeham}},\ }\href {https://doi.org/10.1063/1.3467496}
	{\bibfield  {journal} {\bibinfo  {journal} {J. Phys. Chem. Ref. Data}\
		}\textbf {\bibinfo {volume} {39}},\ \bibinfo {pages} {033105} (\bibinfo
		{year} {2010})}\BibitemShut {NoStop}%
	\bibitem [{\citenamefont {Paradis}\ \emph {et~al.}(2002)\citenamefont
		{Paradis}, \citenamefont {Ishikawa},\ and\ \citenamefont
		{Yoda}}]{Paradis2002}%
	\BibitemOpen
	\bibfield  {author} {\bibinfo {author} {\bibfnamefont {P.-F.}\ \bibnamefont
			{Paradis}}, \bibinfo {author} {\bibfnamefont {T.}~\bibnamefont {Ishikawa}},\
		and\ \bibinfo {author} {\bibfnamefont {S.}~\bibnamefont {Yoda}},\ }\href
	{https://doi.org/10.1023/A:1015459222027} {\bibfield  {journal} {\bibinfo
			{journal} {Int. J. of Thermophys.}\ }\textbf {\bibinfo {volume} {23}},\
		\bibinfo {pages} {825} (\bibinfo {year} {2002})}\BibitemShut {NoStop}%
	\bibitem [{\citenamefont {Ishikawa}\ \emph {et~al.}(2012)\citenamefont
		{Ishikawa}, \citenamefont {Paradis}, \citenamefont {Okada},\ and\
		\citenamefont {Watanabe}}]{Ishikawa2012}%
	\BibitemOpen
	\bibfield  {author} {\bibinfo {author} {\bibfnamefont {T.}~\bibnamefont
			{Ishikawa}}, \bibinfo {author} {\bibfnamefont {P.-F.}\ \bibnamefont
			{Paradis}}, \bibinfo {author} {\bibfnamefont {J.~T.}\ \bibnamefont {Okada}},\
		and\ \bibinfo {author} {\bibfnamefont {Y.}~\bibnamefont {Watanabe}},\ }\href
	{https://doi.org/10.1088/0957-0233/23/2/025305} {\bibfield  {journal}
		{\bibinfo  {journal} {Meas. Sci. Technol.}\ }\textbf {\bibinfo {volume}
			{23}},\ \bibinfo {pages} {025305} (\bibinfo {year} {2012})}\BibitemShut
	{NoStop}%
	\bibitem [{\citenamefont {Behler}(2021)}]{Behler_2021}%
	\BibitemOpen
	\bibfield  {author} {\bibinfo {author} {\bibfnamefont {J.}~\bibnamefont
			{Behler}},\ }\href {https://doi.org/10.1021/acs.chemrev.0c00868} {\bibfield
		{journal} {\bibinfo  {journal} {Chemical Reviews}\ }\textbf {\bibinfo
			{volume} {121}},\ \bibinfo {pages} {10037} (\bibinfo {year} {2021})},\
	\bibinfo {note} {pMID: 33779150},\ \Eprint
	{https://arxiv.org/abs/https://doi.org/10.1021/acs.chemrev.0c00868}
	{https://doi.org/10.1021/acs.chemrev.0c00868} \BibitemShut {NoStop}%
	\bibitem [{\citenamefont {Jakse}\ \emph {et~al.}(2022)\citenamefont {Jakse},
		\citenamefont {Sandberg}, \citenamefont {Granz}, \citenamefont {Saliou},
		\citenamefont {Jarry}, \citenamefont {Devijver}, \citenamefont {Voigtmann},
		\citenamefont {Horbach},\ and\ \citenamefont {Meyer}}]{Jakse2023}%
	\BibitemOpen
	\bibfield  {author} {\bibinfo {author} {\bibfnamefont {N.}~\bibnamefont
			{Jakse}}, \bibinfo {author} {\bibfnamefont {J.}~\bibnamefont {Sandberg}},
		\bibinfo {author} {\bibfnamefont {L.~F.}\ \bibnamefont {Granz}}, \bibinfo
		{author} {\bibfnamefont {A.}~\bibnamefont {Saliou}}, \bibinfo {author}
		{\bibfnamefont {P.}~\bibnamefont {Jarry}}, \bibinfo {author} {\bibfnamefont
			{E.}~\bibnamefont {Devijver}}, \bibinfo {author} {\bibfnamefont
			{T.}~\bibnamefont {Voigtmann}}, \bibinfo {author} {\bibfnamefont
			{J.}~\bibnamefont {Horbach}},\ and\ \bibinfo {author} {\bibfnamefont
			{A.}~\bibnamefont {Meyer}},\ }\href
	{https://doi.org/10.1088/1361-648X/ac9d7d} {\bibfield  {journal} {\bibinfo
			{journal} {J. Phys.: Condensed Matter}\ }\textbf {\bibinfo {volume} {35}},\
		\bibinfo {pages} {035402} (\bibinfo {year} {2022})}\BibitemShut {NoStop}%
\end{thebibliography}
%
\end{document}